\documentclass[aps,prl,twocolumn,showpacs,superscriptaddress,10pt]{revtex4-1}

\usepackage{graphicx}
\usepackage{amsmath,amssymb,amstext,amsthm}
\usepackage{bbold}
\usepackage{cancel}
\usepackage{color}
\usepackage{verbatim}
\usepackage{booktabs}
\usepackage[caption=false]{subfig}

\renewcommand{\Im}{\mathrm{Im}\,}

\newcommand{\Z}{\mathbb{Z}}

\newcommand{\R}{\mathbb{R}}

\newcommand{\id}{\mathbb{1}}

\usepackage[utf8x]{inputenc}
\usepackage{afterpage}

\begin{document}

\title{Generic coexistence of Fermi arcs and Dirac cones \\on the surface of
time-reversal invariant Weyl semimetals}

\author{Alexander Lau}
\affiliation{Institute for Theoretical Solid State Physics, IFW Dresden, 
01171 Dresden, Germany}

\author{Klaus Koepernik}
\affiliation{Institute for Theoretical Solid State Physics, IFW Dresden, 
01171 Dresden, Germany}

\author{Jeroen van den Brink}
\affiliation{Institute for Theoretical Solid State Physics, IFW Dresden, 
01171 Dresden, Germany}
\affiliation{Institute for Theoretical Physics, TU Dresden, 01069 Dresden,
Germany}

\author{Carmine Ortix}
\affiliation{Institute for Theoretical Solid State Physics, IFW Dresden, 
01171 Dresden, Germany}
\affiliation{Institute for Theoretical Physics, Center for Extreme Matter and
Emergent Phenomena, Utrecht University, Princetonplein 5, 3584 CC Utrecht,
Netherlands}

\date{\today}

\begin{abstract}
The hallmark of Weyl semimetals is the existence of open constant-energy contours on their surface -- 
the so-called Fermi arcs -- connecting Weyl points.
Here, we show that for time-reversal symmetric realizations of Weyl semimetals these Fermi arcs in many cases coexist with closed Fermi pockets originating from surface Dirac cones pinned to time-reversal invariant momenta. The existence of Fermi pockets is required for certain Fermi-arc connectivities due to additional restrictions imposed by the six $\Z_2$ topological invariants
characterizing a generic time-reversal invariant Weyl semimetal. We show that a change of the Fermi-arc connectivity generally leads to a different topology of the surface Fermi surface, and identify the half-Heusler compound LaPtBi under in-plane compressive strain as a material that realizes this surface Lifshitz transition.
We also discuss universal features of this coexistence in quasi-particle interference spectra.
\end{abstract}

\maketitle

\paragraph{Introduction -- }

Sparked by the discovery of the quantum Hall effect and its theoretical
explanation~\cite{KDP80,TKN82,Koh85}, the study of topological phases of matter has been one of
the driving forces in modern condensed matter 
physics~\cite{QiZ11,HaK10,SaA16,Bur16,Sen15,LOB15_2}.
Novel states of matter that emerged from these investigations
are, for instance,
time-reversal invariant (TRI) topological 
insulators~\cite{KaM05,FKM07,BHZ06,KWB07,XQH09,PRK15}, and crystalline 
topological insulators~\cite{AnF15,Fu11,TRS12,LBO16},
which are realized in several material systems.
In recent years, the family
of topological materials has been extended by topological
semimetals~\cite{Bur16}. A milestone was the experimental discovery of Weyl semimetals (WSMs)~\cite{HXB15,LWF15,XBA15,HKE16}.
WSMs are three-dimensional gapless materials whose bulk energy bands
cross linearly at isolated points, the so-called Weyl nodes, in the Brillouin zone (BZ)~\cite{WTV11,BuB11,ZWB12,Oja13,SGW15}. 
Weyl nodes are characterized by their chirality
and can only be annihilated pairwise. For this reason, they are a robust bulk
feature~\cite{BuB11,OkM14}: generic perturbations 
shift the nodes 
in energy and momentum space
without annihilating them.
Most importantly, WSMs host robust surface states commonly referred to as
Fermi arcs. They form an open Fermi surface connecting 
the surface projections of Weyl nodes with opposite chiralities~\cite{WTV11}.

Recently, it has been theoretically proposed
that Fermi arcs can, under certain conditions, coexist with surface Dirac cones at
the interface
between a time-reversal broken WSM and a three-dimensional (3D) TRI topological insulator~\cite{GVB15,JuT17}.
The close relation between Dirac cones and Fermi arcs is also manifest in the fact that Dirac cones can be created by fusing Weyl points in a specific manner~\cite{OkM14}.
In this Letter, we show that in \emph{time-reversal symmetric} WSMs
the coexistence of Dirac cones and Fermi arcs arises naturally.
It is due to additional restrictions on the surface
Fermi surface imposed by the $\Z_2$ invariants associated with the TRI planes of the WSM.
We thereby show that this coexistence is 
encountered
in the half-Heusler compound LaPtBi, which realizes a Weyl semimetal phase under in-plane compressive strain \cite{RJY16}. Finally, we determine universal features of this coexistence in quasiparticle-interference (QPI)
spectra, which are relevant for scanning tunneling experiments~\cite{DML15,KLW16,CXZ16,BMA16,ZXB16}. 

\paragraph{$\Z_2$ invariants in TRI Weyl semimetals -- }

\begin{figure}[t]\centering
\includegraphics[width=1.0\columnwidth]{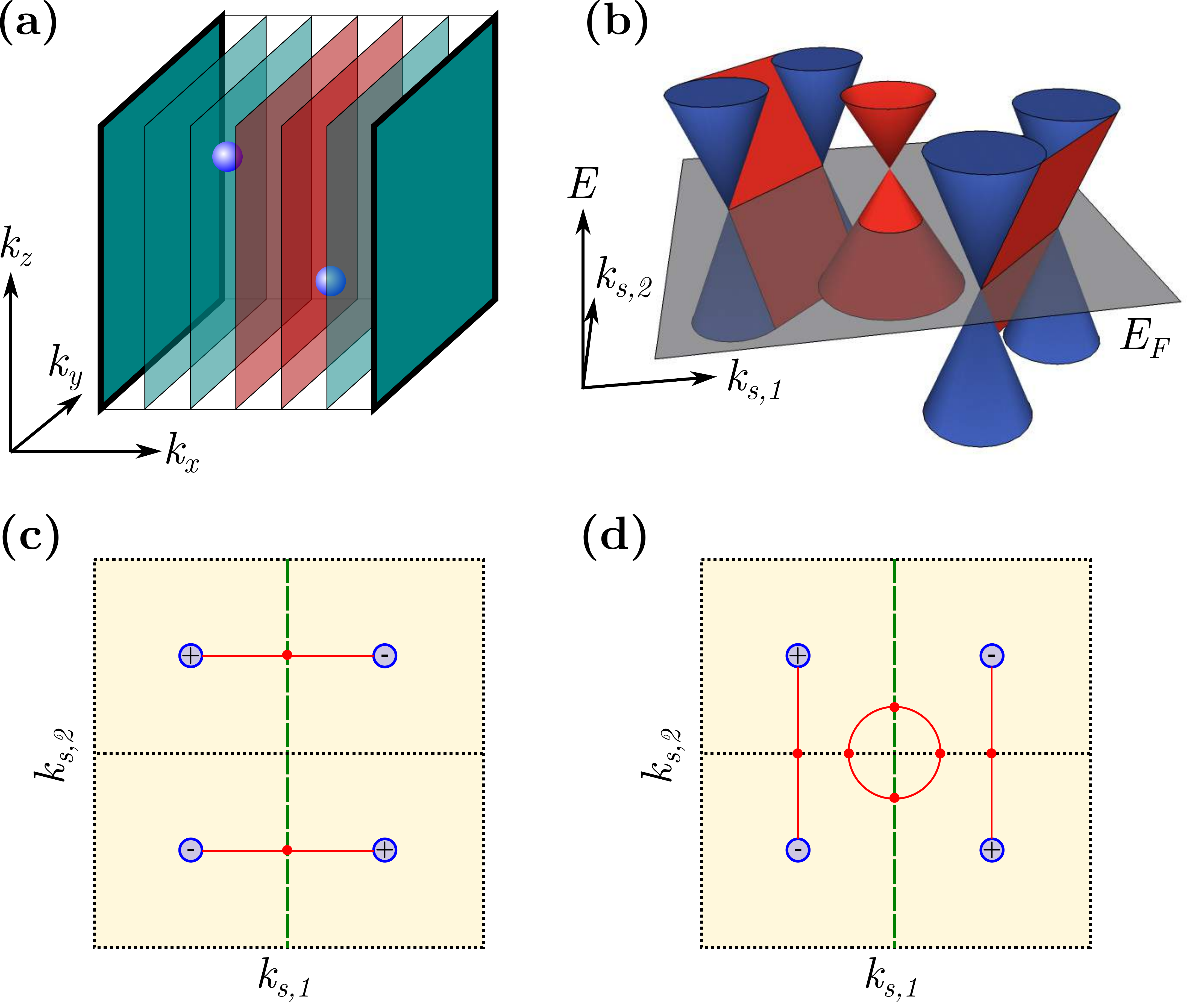}
\caption{(color online) (a) The BZ of a WSM 
as a collection of 2D insulators with zero (green) or nonzero (red) Chern numbers.
Weyl nodes (blue spheres) separate planes with different Chern numbers. 
The bold frames indicate the TRI 2D insulators characterized by a $\Z_2$ invariant.
(b) Typical low-energy surface spectrum of a
TRI WSM with an additional surface Dirac cone: 
surface states are shown in red, whereas the surface projections of the 3D bulk Weyl cones are highlighted in blue.
(c)-(d) Fermi arc connectivities in the surface BZ of a
TRI WSM with four Weyl points indicated by their topological
charge $\pm$. The surface projections of the 3D TRI planes are highlighted 
by dotted black ($\nu=0$) or dashed green ($\nu=1$) lines.}
\label{fig:conceptional_figures}
\end{figure}

It is well known that the existence of Weyl nodes in momentum space requires either time-reversal $\Theta$ or inversion $\cal{I}$ symmetry breaking~\cite{BuB11,ZWB12,OkM14}.
In the presence of both symmetries 
all energy bands are at least doubly degenerate, 
which requires
a linear crossing to be a four-fold degenerate Dirac point. $\Theta$ or $\cal{I}$ symmetry-breaking perturbations split a Dirac point into two separate Weyl points of opposite chirality. 
A Weyl node represents a monopole of the Berry flux $\mathcal{A}(\mathbf{k})$ in momentum space~\cite{WTV11,OkM14}.
Consequently, an integral of the
Berry flux over a closed surface enclosing the Weyl node results in a nonzero integer value, which defines the topological charge or
chirality of the Weyl node~\cite{WTV11,OkM14}. Since the total topological charge in the whole BZ must vanish~\cite{NiN81_2}, Weyl nodes always appear in pairs of opposite chirality. 
Moreover, in TRI systems each Weyl point has a time-reversal partner of same chirality,
which implies
the total number of Weyl nodes in these systems to be $4 n$, with $n$ being an integer~\cite{OkM14}.  
The nonzero topological charge of the Weyl nodes can also be interpreted~\cite{BuB11} as the change in the Chern number of the collection of gapped two-dimensional (2D) systems realized by decomposing the 3D BZ of a WSM in 2D momentum space cuts separating the Weyl points from each other [see Fig.~\ref{fig:conceptional_figures}(a)]. 
This property is at the basis of the existence of one of the most interesting hallmarks of WSMs: the existence of open constant-energy contours in the surface BZ called Fermi arcs~\cite{WTV11}. 

In general, time-reversal invariant WSMs can be additionally characterized by six $\Z_2$ invariants~\cite{CTS16}, except when the Weyl points are pinned to time-reversal invariant momenta, which can only occur in materials with chiral space groups~\cite{CSW16}. The Chern number of the effective 2D insulators realized by the TRI planes will be zero, but the time-reversal polarizations still allow to characterize the effective 2D systems in terms of a $\Z_2$ topological invariant $\nu$~\cite{FuK06}. Contrary to TRI insulators in 3D, where the six $\nu_i$ are not independent and can be reduced to four using homotopy arguments~\cite{MoB07,Roy09} -- the well-known strong and weak indices of 3D TRI 
insulators~\cite{FuK06,FKM07} -- in a time-reversal invariant WSM all six $\Z_2$ invariants are independent and characterize the WSM as explained below. 

For a generic surface of a WSM, by bulk-boundary correspondence the $\nu_i$ determine whether an even  
($\nu_i=0$) or odd ($\nu_i=1$) number of Kramers pairs of surface states cross the Fermi level along the surface projection of the $i$-th TRI plane. This imposes restrictions on the structure of the surface Fermi surface but still does not uniquely determine it. Figs.~\ref{fig:conceptional_figures}(c) and~(d) sketch two 
allowed but qualitatively very different
surface Fermi surfaces of a time-reversal invariant WSM in which the $\Z_2$ invariants of the planes $k_{s,2}=0,\pi$ and $k_{s,1}=\pi$ have the trivial value $0$, whereas the invariant associated with the plane at $k_{s,1}=0$ has the nontrivial value $1$. A surface Fermi surface consisting of only two open arcs, connecting Weyl points as depicted in Fig.~\ref{fig:conceptional_figures}(c), is entirely allowed. 
However, different pairs of Weyl points of opposite chirality can be connected only if an additional Fermi pocket, enclosing a time-reversal invariant point, is created [see Fig.~\ref{fig:conceptional_figures}(d)]. The latter situation is a unique signature of Fermi arcs coexisting with a 
surface Dirac cone [see Fig.~\ref{fig:conceptional_figures}(b) and Ref.~\onlinecite{supp}], which is an exclusive feature of TRI Weyl semimetals. This surface Dirac cone is protected for a given connectivity of the Fermi arcs since its corresponding
Fermi pocket can only be removed by connecting it to the Fermi arcs, which would lead to another reconnection of the Weyl nodes. We emphasize that while 
this transition does not change the $\Z_2$ invariants of the time-reversal invariant
WSM, the change of the Fermi surface topology does imply a Lifshitz
transition on the surface of the material~\cite{Vol16,VMB15}.
Before studying this Lifshitz transition in an explicit Hamiltonian, we first determine the
generic consequences for QPI. This we compare to explicit QPI calculations for the explicit Hamiltonian later on.

\paragraph{Phenomenological QPI patterns -- } Having established the coexistence of Fermi arcs and Dirac cones in time-reversal invariant WSM, we now proceed to analyze their fingerprints in QPI patterns, which can be observed in scanning tunneling spectroscopy experiments~\cite{DML15,KLW16,CXZ16,BMA16,ZXB16}.
We start out by noticing that QPI spectra can be approximated in terms of the  
joint density of states (JDOS)~\cite{SVA07,DML15},
\begin{equation}
J(\mathbf{q},E) = \int d^2 k\: A(\mathbf{k}+\mathbf{q},E)A(\mathbf{k},E),
\end{equation}
where $\mathbf{k}$ is the momentum parallel to the surface, and 
$A(\mathbf{k},E)=-1/2\pi\,\mathrm{Im}\lbrace \mathrm{Tr}[G_s(\mathbf{k},E)]\rbrace$
is the spectral function with the surface Green's function $G_s(\mathbf{k},E)$.

To understand the characteristic features arising in the JDOS, we have performed a phenomological 
analysis based on a simple ansatz for the spectral function. We present the details of our calculation in the Supplemental Material~\cite{supp} and proceed here with the presentation of the key results. First of all, our analysis reproduces the results of previous studies, namely a pinch point at $\mathbf{q} \equiv 0$~\cite{KLW16} surrounded by crescent-shaped patterns due to the contribution of the Fermi arcs~\cite{MiF16}, and a
disk feature associated with the presence of a Dirac cone~\cite{GuF10}. Most importantly, for the case of coexistence we find 
two additional kidney-shaped features at a distance $|\mathbf{q}|$ corresponding to the distance between the Fermi arcs and the Fermi pocket. Furthermore, the broadening of these patterns corresponds to the diameter of the Fermi pocket. We can unambiguously attribute this feature to scattering events between Fermi pocket and Fermi arcs.
Hence, they represent the universal QPI feature of the coexistence of Fermi arcs with a Dirac cone 
on the surface of time-reversal invariant WSMs.

\paragraph{Tight-binding model formulation -- } 
Next, we introduce a tight-binding model for a time-reversal invariant WSM to investigate on a microscopic basis the coexistence of surface Dirac cones and Fermi arcs.
The tight-binding model is defined on a cubic lattice and reads
\begin{eqnarray}
H(\mathbf{k}) &=& a\,(\sin k_x \,\tau^1 s^3
+ \sin k_y\, \tau^2 s^0) + \beta\, \tau^2 s^2 + d\, \tau^2 s^3 \nonumber\\
&&{} + [t\cos k_z + 2b(2 - \cos k_x - \cos k_y)]\,\tau^3 s^0 \nonumber\\
&&{}+ \alpha\sin k_y\, \tau^1 s^2 + \lambda \sin k_z\, \tau^0 s^1,
\label{eq:tb_Hamiltonian}
\end{eqnarray}
where the $s^i$ are Pauli matrices in spin space, whereas the $\tau^i$
are Pauli matrices associated with additional orbital degrees of freedom. The
lattice constant has been set to unity. The Hamiltonian is based on 
a general tight-binding model introduced in Ref.~\onlinecite{KLW16}.
The model preserves time-reversal symmetry with $\Theta = i\tau^0
s^2\,K$, $\mathbf{k}\rightarrow -\mathbf{k}$, where $K$ is complex
conjugation. The $\beta$, $d$ and $\lambda$ terms break inversion symmetry
with the inversion operator $P=\tau^3 s^0$, $\mathbf{k}\rightarrow
-\mathbf{k}$, which is a necessary condition for the existence of a Weyl
semimetal phase.

\begin{figure}[t]\centering
\includegraphics[width=0.9\columnwidth]
{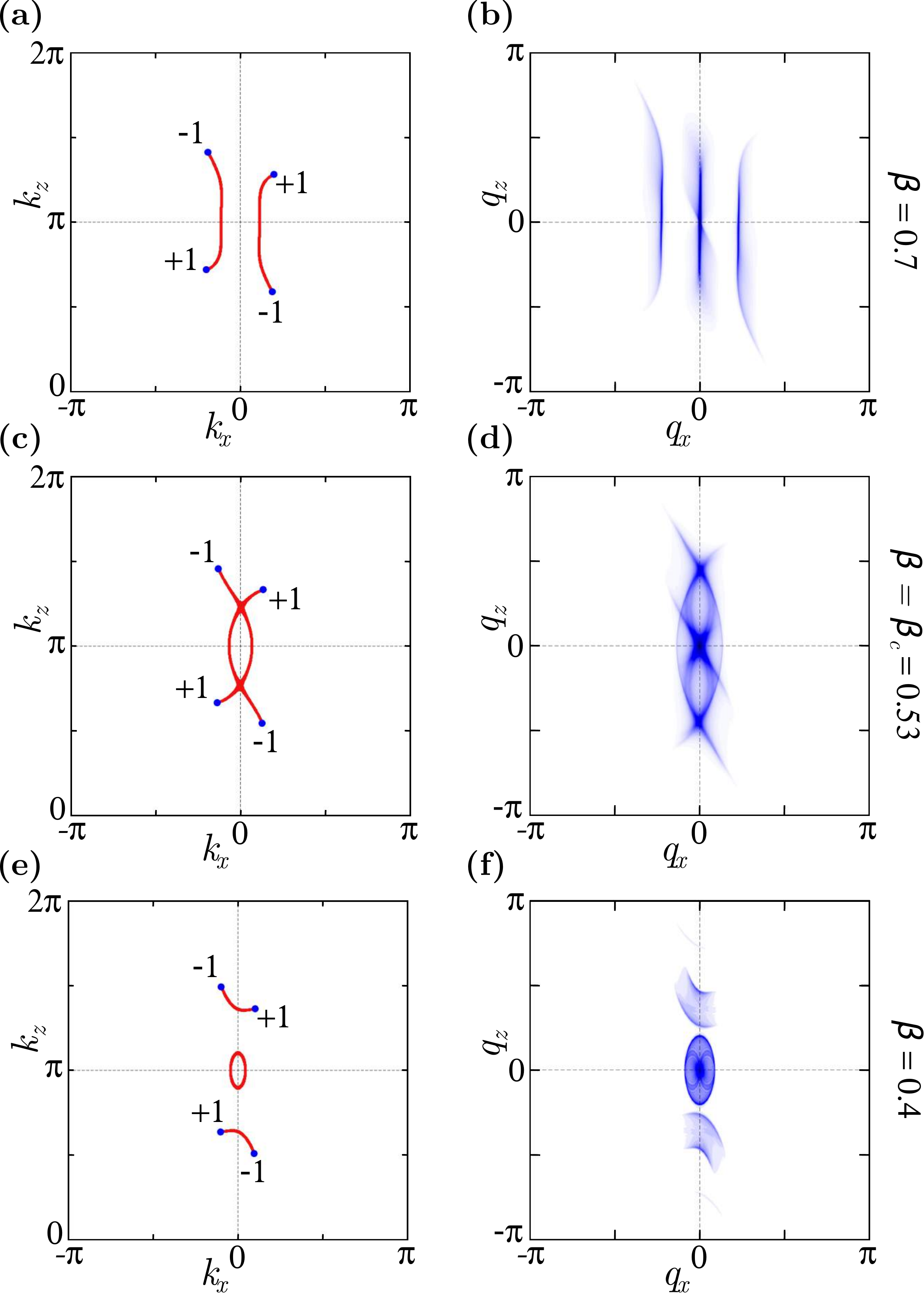}
\caption{(color online) Fermi surfaces and JDOS for (010) surfaces in the tight-binding model with $a=b=1$, $t=1.5$, $\alpha=0.3$, $d=0.1$, $\lambda=0.5$, and $E_F=0$: the first column shows the Fermi surfaces for different values of the parameter $\beta$. The bulk Weyl nodes are highlighted in blue and their topological charge is indicated. Surface states are highlighted in red. The second column shows the corresponding JDOS spectra. In (f), the kidney-shaped features indicative of the coexistence of Fermi arcs and Dirac cones are clearly visible in the JDOS.}
\label{fig:tm_model_figures}
\end{figure} 
To demonstrate the coexistence of surface Dirac cones and Fermi arcs in $H(\mathbf{k})$,
we start from a particular Weyl semimetal phase and vary the parameter $\beta$. 
The results are presented in Fig.~\ref{fig:tm_model_figures}.
With the chosen parameters, the model features four bulk Weyl points
with topological charge $\pm 1$, as can be determined by integrating the Berry flux of the bulk Hamiltonian in Eq.~\eqref{eq:tb_Hamiltonian} over the surface of a BZ volume containing the Weyl node.
The fact that the Weyl nodes are all located away from the TRI planes of the bulk BZ also allows one to calculate all the six $\Z_2$ invariants using their Wannier-center formulation (see Refs.~\onlinecite{supp} and~\onlinecite{YQB11}).
In particular, we find that $\nu_{k_z=\pi}=1$ 
while the remaining five $\Z_2$ invariants are zero. 
At the (010) surface we therefore expect an odd number of Kramers pairs at $k_z=\pi$ and an even number of Kramers pairs at $k_z=0$ and $k_x=0,\pi$. 
\begin{center}
\begin{figure*}[tbp]
\includegraphics[width=1.0\textwidth]
{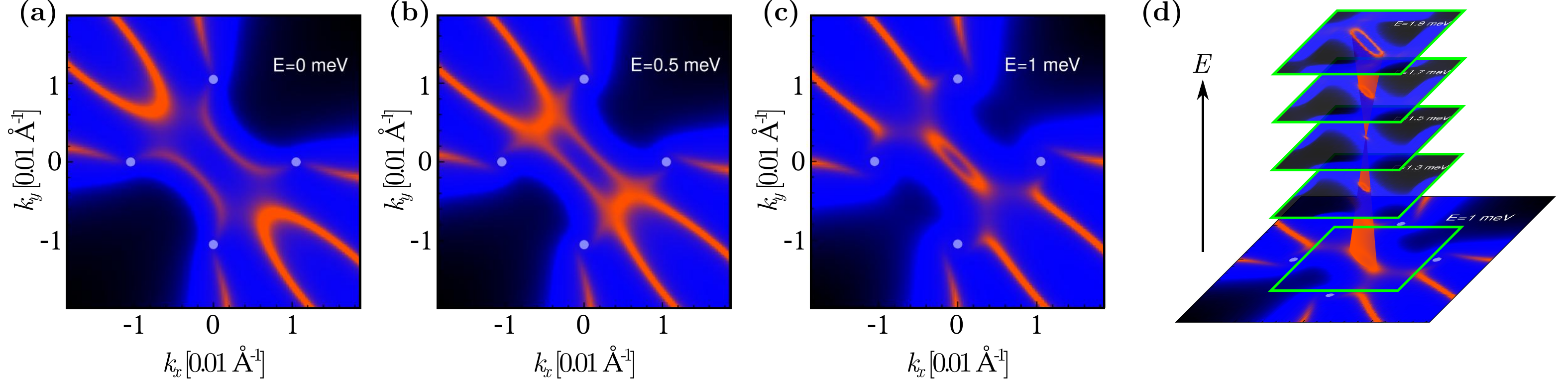}
\caption{(color online) Surface Fermi surfaces of LaPtBi with (001) termination: shown is the surface spectral weight. The positions of the four Weyl-point projections are marked by grey dots. The panels display the transition between different Fermi-arc connectivities by varying the Fermi level. Note that the connectivity shown in (c) requires the presence of an additional Fermi pocket around the origin which resembles Fig.~\ref{fig:tm_model_figures}(e). (d) A further increase of the Fermi level reveals that the Fermi pocket indeed originates from a Dirac cone around $\bar{\Gamma}$.}
\label{fig:LaPtBi_BZ_zoom}
\end{figure*}
\end{center} 
For large values of $\beta$, we find that Fermi arcs connect two Weyl 
nodes in the left half-plane and two Weyl nodes in the right half-plane [see Fig.~\ref{fig:tm_model_figures}(a)]. The fact that the Fermi arcs cross only the line $k_z=\pi$ is in agreement with the
values of the topological invariants. In Fig.~\ref{fig:tm_model_figures}(b) we show the corresponding JDOS based on the (010) surface Green's function of the system (see Refs.~\onlinecite{supp} and~\onlinecite{SSR85} for computational details).
We see a pinch point surrounded by two crescent-shaped patterns in agreement with our general consideration.

By decreasing the parameter $\beta$, the Fermi arcs are bent towards each other until they intersect at a critical value of $\beta$ [see Fig.~\ref{fig:tm_model_figures}(c)]. 
This point corresponds to the Lifshitz transition at which the connectivity of the Fermi arcs changes.
A further decrease in the parameter $\beta$ reveals the coexistence of Dirac cones 
and Fermi arcs [see Fig.~\ref{fig:tm_model_figures}(e)]: 
surface Fermi arcs 
connect two Weyl nodes in the upper half-plane and two Weyl
nodes in the lower half-plane of the surface BZ. In addition
to the open-arc features, we find an elliptical Fermi pocket of surface states
around the $\bar{Z}$ point of the surface BZ. The existence of the Fermi pocket is
required for this particular connectivity of Weyl nodes to satisfy the number
of surface states imposed by the topological invariants $\nu_i$ which have not changed during the Lifshitz transition. From an analysis of the surface band structure~\cite{supp}, we find that the Fermi pocket originates from a \emph{dangling} surface Dirac cone. This type of Dirac cone connects bulk conduction to bulk valence bands not along all directions of the surface BZ, as would be the case for a 3D topological insulator. This is only possible in TRI Weyl semimetals due to the lack of a global relation between the $\Z_2$ invariants.

The JDOS of this configuration is in perfect agreement with our analytical considerations (see Ref.~\onlinecite{supp}): we find a new elliptical feature around the origin associated with the additional Fermi pocket. The characteristic crescent-shaped Fermi-arc features are barely visible due to the small size of the Fermi arcs. Most importantly, we find the kidney-shaped features indicative of scattering between the Fermi arcs and the Fermi pocket. Their broadening equals the size of the Fermi pocket whereas their position corresponds to the distance between Fermi pocket and Fermi arcs. A comparison of Figs.~\ref{fig:tm_model_figures}(b) and~(f) shows that the different Fermi-arc connectivities are reflected in distinct universal JDOS features.

\paragraph{Strained LaPtBi -- }
Having established the coexistence of Dirac cones and Fermi arcs in a generic tight-binding model of TRI Weyl semimetals, we next show its realization in the half-Heusler compound LaPtBi, which possesses both band inversion \cite{XYF10} and lattice noncentrosymmetry. 
A recent theoretical ab-initio study suggests that
under a broad range of in-plane biaxial compressive strain, LaPtBi realizes a Weyl semimetal phase with eight Weyl nodes residing precisely at the Fermi level at stoichiometry composition~\cite{RJY16}.
We have performed density functional theory calculations employing the Full Potential
Local Orbital (FPLO) method~\cite{KoE99} in 4-component relativisitic mode (see Ref.~\onlinecite{supp} for computational details), applying a compressive in-plane strain of $a=0.99a_{0}$, $c=1.02a_{0}$.
For the study of surface states, we investigate a semi-infinite slab 
with a (001) surface corresponding to a termination along one of the LaBi planes.

Due to the concomitant presence of time-reversal and two-fold rotations along the $x$ and $y$ axis, eight Weyl points of charge $\pm 1$ are located at the $k_x=0$ and $k_y=0$ planes of the bulk BZ. 
Moreover, the electronic states in the $k_x=k_y$ and $k_x=-k_y$ planes are all gapped, so that the $\Z_2$ topological invariant in these TRI subsystems is well-defined.
In the (001) surface BZ, the Weyl points are projected pairwise on
four different surface momenta thereby giving the projected Weyl points an effective topological charge of $\pm 2$. Hence, there
must be two outgoing Fermi arcs for each Weyl-point
projection. Moreover, we find that the projections of the TRI planes
$k_x=k_y$ and $k_x=-k_y$ feature an odd number of
surface Kramers pairs (see Fig.~\ref{fig:LaPtBi_BZ_zoom}, and
Fig.~2 in Ref.~\onlinecite{supp}). 
This implies non-trivial $\Z_2$ invariants which we have confirmed by explicit calculations using the Wannier-center formulation of the topological number (see Ref.~\onlinecite{supp}).
This gives rise to restrictions on the Fermi surface topology. In particular, we find that we can tune between different Fermi-arc connectivities by varying the Fermi level (see Fig.~\ref{fig:LaPtBi_BZ_zoom}), which can be accomplished for instance by doping.

In Fig.~\ref{fig:LaPtBi_BZ_zoom}(a), the Fermi level coincides with the Weyl-point energies. In this case, the Fermi arcs connect in a way that does not require an additional Fermi pocket. By raising the Fermi level, a Lifshitz transition takes place [compare Fig.~\ref{fig:LaPtBi_BZ_zoom}(b) to Fig.~\ref{fig:tm_model_figures}(c)]. Finally, the connectivity of the Weyl nodes switches which leads to the emergence of an additional Fermi pocket around the projected $\Gamma$ point, as shown in Fig.~\ref{fig:LaPtBi_BZ_zoom}(c). This Fermi pocket is indeed associated with a surface Dirac cone [see Fig.~\ref{fig:LaPtBi_BZ_zoom}(d)] as one can infer from surface Fermi surfaces at larger $E_F$ and from the dispersion along high-symmetry cuts through the surface BZ 
(see Ref.~\onlinecite{supp}).

\paragraph{Conclusions -- }

Besides the topological charges of the Weyl nodes, a generic TRI Weyl semimetal can be characterized by six $\Z_2$ invariants associated with the TRI planes of the 3D BZ.
For the surface of a time-reversal invariant WSM, they impose restrictions on the number of surface Kramers pairs along the surface projections of the TRI planes and, therefore, also on the structure of the Fermi arcs which connect the Weyl nodes on the surface. Nevertheless, a remaining modulo-two ambiguity gives rise to many possible and qualitatively different Fermi-arc connectivities. In particular, certain connectivities require the creation of a Fermi pocket which is connected to the presence of a surface Dirac cone pinned to a TRI momentum. This changes the topology of the Fermi surface and is, thus, accompanied by a Lifshitz transition. It is crucial to note that this transition does not change the $\Z_2$ invariants and is, therefore, generic to all TRI Weyl semimetals.
We have further shown that the coexistence of Fermi arcs and Dirac cones leads to universal, kidney-shaped features in QPI patterns which are accessible in scanning tunneling spectroscopy experiments.
Finally, our density-functional theory calculations show that the half-Heusler compound LaPtBi under compressive strain is a good candidate material to realize the general coexistence established in this work.

We acknowledge the financial support of the Future and Emerging Technologies (FET) programme within
the Seventh Framework Programme for Research of the European Commission 
under FET-Open grant number: 618083 (CNTQC).  This work has been 
supported by the Deutsche Forschungsgemeinschaft under Grant No. OR 404/1-1 and
SFB 1143. C.O. acknowledges support from a VIDI grant (Project 680-47-543) financed by the Netherlands Organization for Scientific Research (NWO).

\clearpage

\newpage

%%%%%%%%%%%%%%%%%%%%%%%%%%%%%%%%%%%%%%%%%%%%%%%%%%%%%%%%%%%%%%%%%%%%%%%%%%%%%%%%%%%%%%%%%%%%%%%
%%%%%%%%%%%%%%%%%%%%%%%%%%%%%%%%% SUPPLEMENTAL MATERIAL %%%%%%%%%%%%%%%%%%%%%%%%%%%%%%%%%%%%%%%
%%%%%%%%%%%%%%%%%%%%%%%%%%%%%%%%%%%%%%%%%%%%%%%%%%%%%%%%%%%%%%%%%%%%%%%%%%%%%%%%%%%%%%%%%%%%%%%

\section*{SUPPLEMENTAL MATERIAL}

\setcounter{figure}{0}    

\subsection*{A: Density-functional theory calculations for LaPtBi}

We have performed density functional theory (DFT) calculations to investigate
the half-Heusler compound LaPtBi under compressive in-plane strain. Our investigation is motivated
by recent theoretical results shown in Ref.~\onlinecite{RJY16} in which LaPtBi is predicted to realize a Weyl semimetal phase.
LaPtBi has a noncentrosymmetric 
crystal structure with a lattice constant of $a_0=6.829\textrm{\AA}$~\cite{RJY16}. Furthermore, it preserves time-reversal symmetry which makes it a suitable material for our study.

For the DFT calculations, the Full Potential
Local Orbital method~\cite{KoE99} in 4-component relativisitic mode
was employed, using $12^{3}$ points for the tetrahedron integration
method. The resulting bands were fitted via maximally projected Wannier
functions including a minimum basis of La 6s6p5d4f, Bi 6s6p6d and
Pt 6s6p5d. The Wannier model is used to prove the existence of the
Weyl points via Berry curvature monopoles and is mapped onto a semi
infinite slab with LaBi-plane (001) termination for the determination of
the surface spectral function.

\begin{figure}[t]\centering
\includegraphics[width=1.0\columnwidth]
{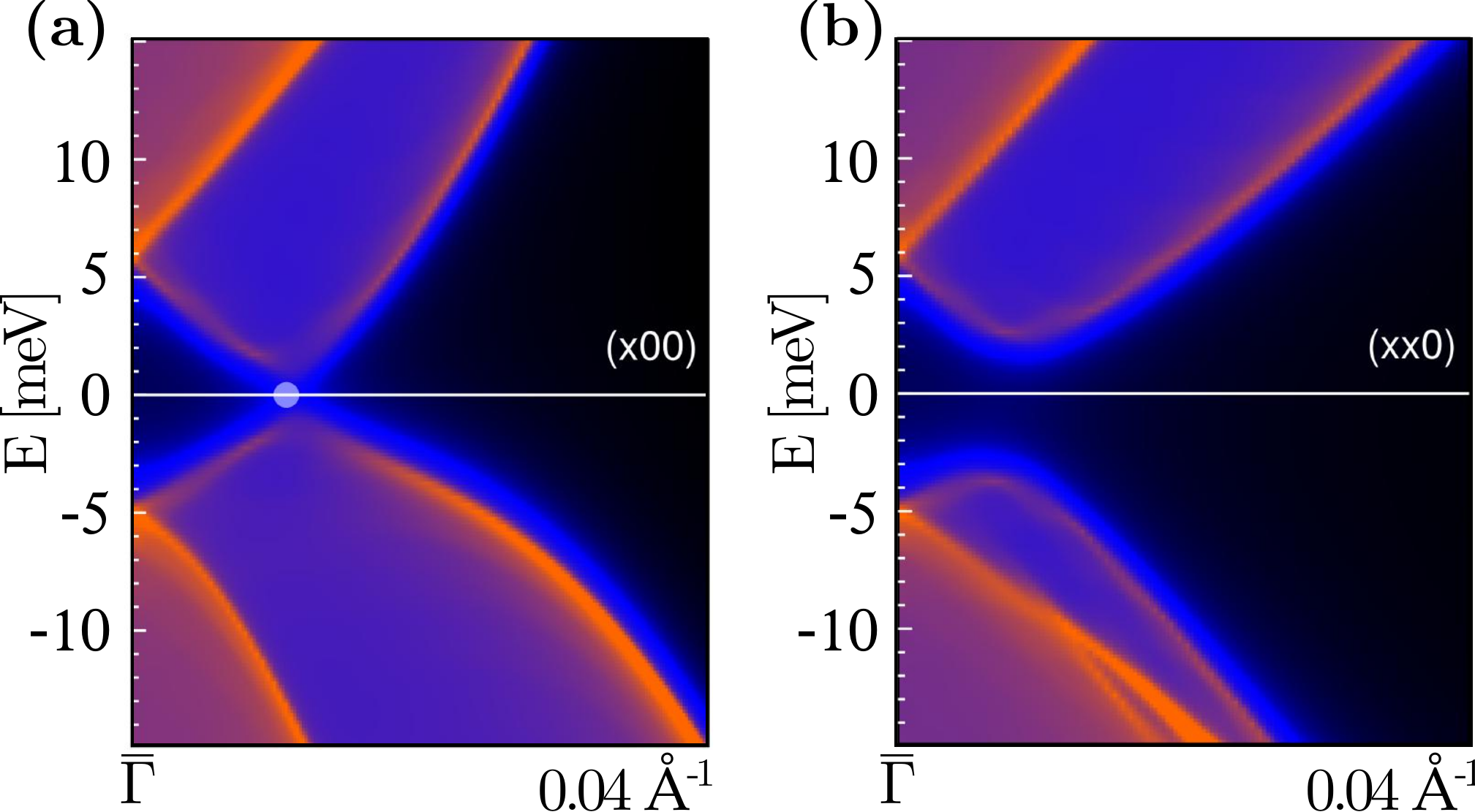}
\caption{(color online) Bulk energy bands of LaPtBi projected onto the (001) surface BZ along high-symmetry directions close to the $\bar{\Gamma}$ point: 
(a) $\left(x00\right)$ direction (surface projection of $k_y=0$ plane). (b) $\left(xx0\right)$ direction (surface projection of $k_x=k_y$ plane). Weyl node projections are marked by grey dots.}
\label{fig:LaPtBi_bulk_BZ}
\end{figure}

We applied compressive in-plane strain of $a=0.99a_{0}$, $c=1.02a_{0}$
following Ref.~\onlinecite{RJY16}. We find eight Weyl points at positions
$(\pm k_{x}^*,0,k_{z}^*)$ and $(0,\pm k_{y}^*,k_{z}^*)$, with $k_{x}^*=k_{y}^*=0.0106\textrm{\AA}^{-1}$ and 
$k_{z}^{*}=0.041\textrm{\AA}^{-1}$, which agrees very well with the results in Ref.~\onlinecite{RJY16}.
In Fig.~\ref{fig:LaPtBi_bulk_BZ}, we show the bulk bands projected
onto the (001) surface BZ along the $\left(x00\right)$ and $\left(xx0\right)$
direction close to the $\bar{\Gamma}$ point. More specifically, the shown slices correspond to surface projections of the $k_y=0$ and $k_y=k_x$ planes of the bulk BZ. As we can see, there is a bulk gap-closing point at $E=0$ along $k_y=0$ [see Fig.~\ref{fig:LaPtBi_bulk_BZ}(a)], which corresponds to the projection of two Weyl nodes, whereas the system has a full bulk gap along the $k_y=k_x$ plane [see Fig.~\ref{fig:LaPtBi_bulk_BZ}(b)]. Similar observations are made for the planes with $k_x=0$ and $k_y=-k_x$, respectively (not shown).

\begin{figure}[t]\centering
\includegraphics[width=1.0\columnwidth]
{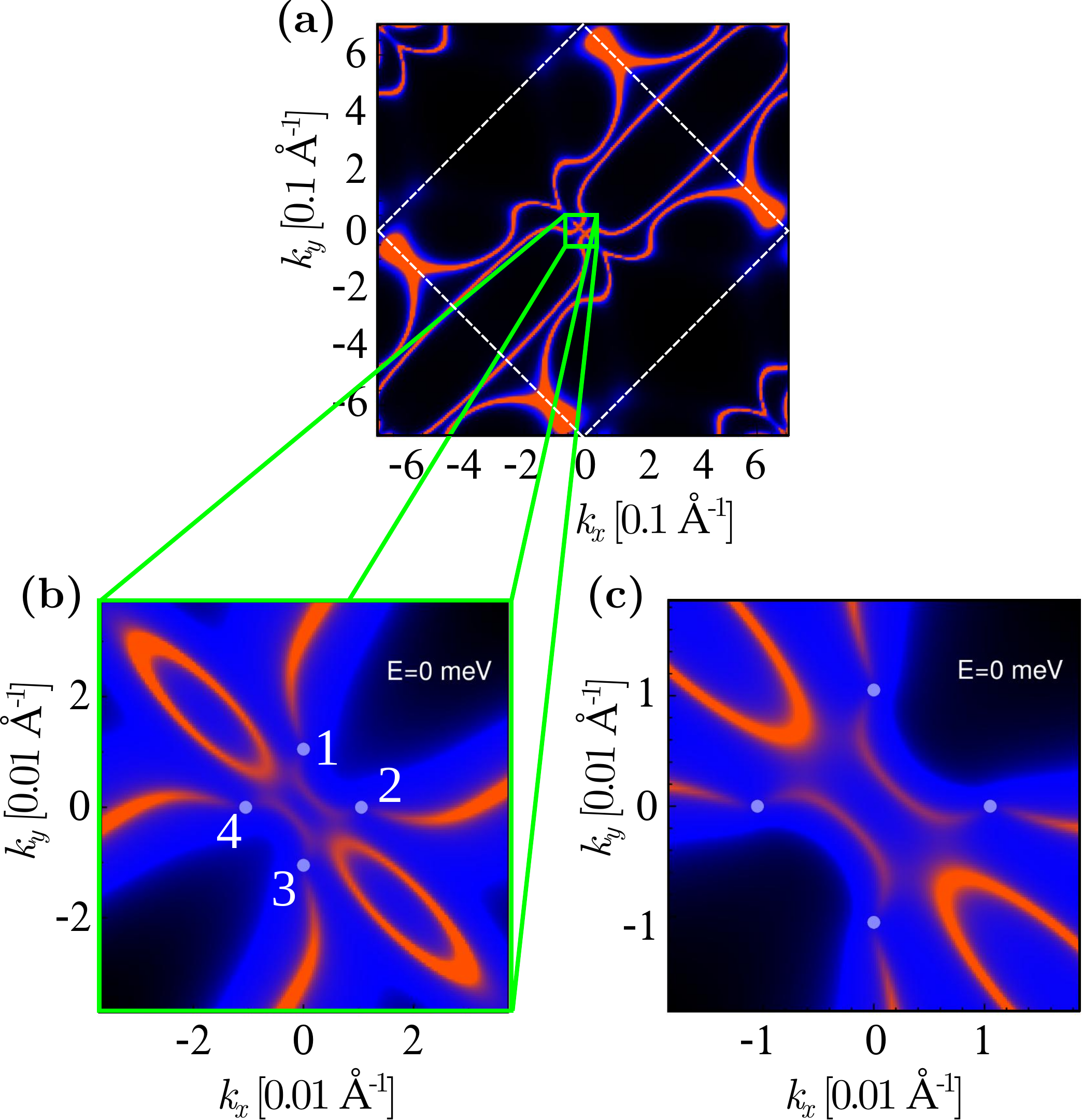}
\caption{(color online) Surface spectral weight in the (001) surface BZ of LaPtBi for $E=0$: (a) entire surface BZ. Borders of the surface BZ are indicated by white dashed lines. The region inside the green square is magnified in panel (b). (b) Magnified region around the $\bar{\Gamma}$ point of the surface BZ. The positions of the projected Weyl nodes are marked by grey dots. For clarity, the Weyl node projections have been labeled by numbers. Panel (c) shows the same region with a higher magnification factor. Note how the surface Fermi arcs connect the Weyl node projections.}
\label{fig:LaPtBi_BZ_several_zooms}
\end{figure} 
Fig.~\ref{fig:LaPtBi_BZ_several_zooms} shows our result for the surface
spectral weight in the $k_{x}k_{y}$-plane of the LaBi terminated
$\left(001\right)$ surface of a semi infinite slab. In panel~(a), we display 
the full surface Fermi surface of the system in the (001) surface BZ, whereas panels~(b) and~(c)
show a magnified region around the $\bar{\Gamma}$ point with different zoom factors. The four Weyl point projections along the $k_x=0$ and $k_y=0$ lines are highlighted in grey and labeled by numbers for clarity.
We note that each point corresponds to the projection of two Weyl nodes with the same charge giving each point an effective Weyl charge of $\pm 2$. 

In agreement with the projected charge, we find two outgoing lines for each Weyl point projection which correspond to the Fermi arcs. From the shown figures we infer the following Fermi arc connectivity:  node 1 is connected to node 2, node 2 is connected to node 3 through the border of the surface BZ, node 3 is connected to node 4, and node 4 is connected to node 1 again through the border of the surface BZ. This means that all Weyl node projections lie along \emph{one} closed loop of Fermi arcs. Furthermore, along $k_x=-k_y$ we find two trivial Fermi pockets and there are also two more trivial lines of surface states further away from the origin.

Let us now analyze the surface Fermi surface in the light of the results presented in the main part of the Letter. As demonstrated above, the time-reversal invariant planes $k_x=k_y$ and $k_x=-k_y$ have full energy gaps in the bulk. Therefore, they represent 2D TRI insulators. The corresponding $\Z_2$ invariant can be inferred by bulk-boundary correspondence by simply counting the number of Kramers pairs in the (001) surface BZ. For the $k_x=k_y$ plane we count \emph{one} Kramers pair of surface states and \emph{three} for the $k_x=-k_y$ plane. Hence, both planes represent nontrivial 2D insulators with $\Z_2$ invariant $\nu=1$.
We have confirmed this by computing the corresponding invariants explicitly using the Wannier-center formulation of the $\Z_2$ invariant discussed in Sec.~C of this Supplemental Material.

\begin{figure}[t]\centering
\includegraphics[width=1.0\columnwidth]
{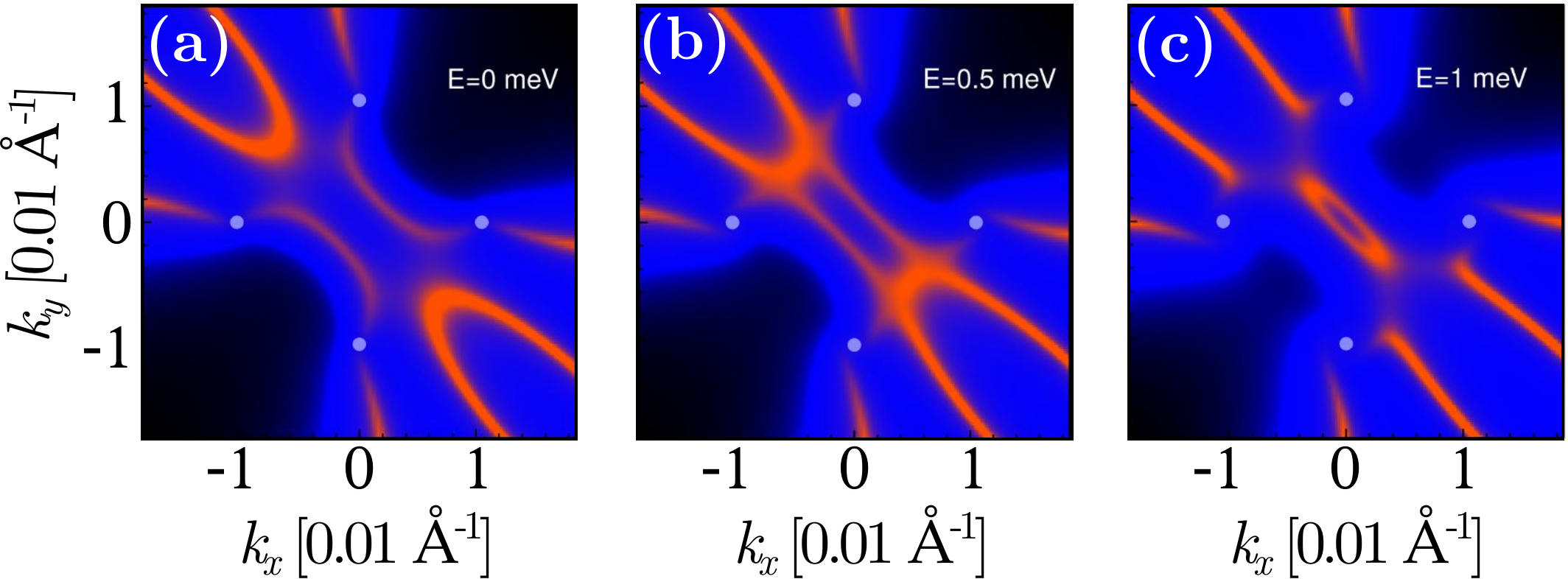}
\caption{(color online) Surface spectral weight in the (001) surface BZ of LaPtBi with varying Fermi level: (a) $E=0\,\textrm{meV}$, (b) $E=0.5\,\textrm{meV}$, (c) $E=1.0\,\textrm{meV}$. The positions of the Weyl node projections are marked by grey dots. Note how the Fermi arcs reconnect in (b) and an additional Fermi pocket is created around the origin in (c).}
\label{fig:LaPtBi_zoom_supp}
\end{figure}

Let us now study what happens if we increase the Fermi level, which could for instance be achieved by electron doping. Our results are shown in Fig.~\ref{fig:LaPtBi_zoom_supp}. We observe that the Fermi arcs connecting node 1 to node 2 and node 3 to node 4 start to fuse with the trivial Fermi pockets [see Fig.~\ref{fig:LaPtBi_zoom_supp}(b)]. At this point the Weyl nodes reconnect. Starting from the given Fermi arc connectivity, such a reconnection is only possible by creating an additional Fermi pocket around the origin in order to respect the constraints imposed by the $\Z_2$ invariants of the TRI planes. This is shown in Fig.~\ref{fig:LaPtBi_zoom_supp}(c).
Note that the Weyl point projections are indeed connected in a different way. More specifically, node 1 and node 4 are now \emph{directly} connected by two Fermi arcs. The same holds for node 2 and node 3. This means that the Fermi arcs now form \emph{two} closed loops each including two Weyl node projection points.

\begin{figure}[t]\centering
\includegraphics[width=1.0\columnwidth]
{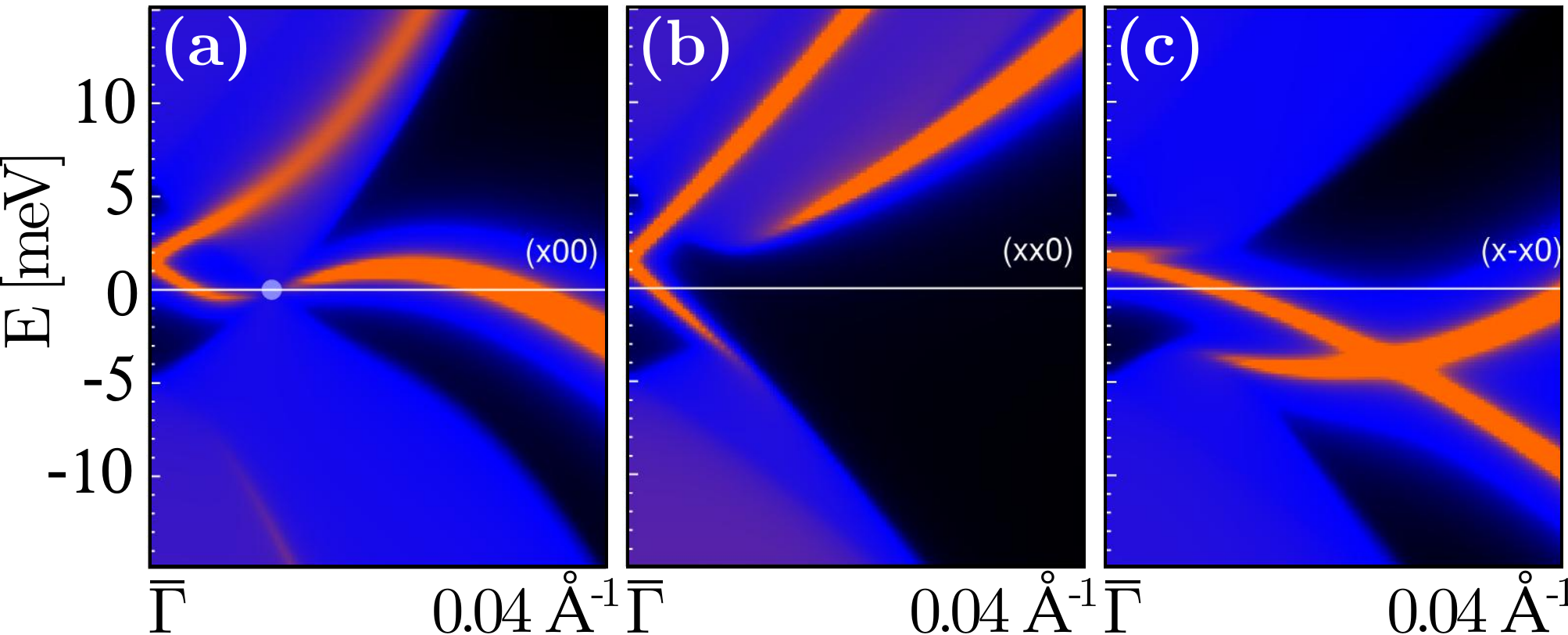}
\caption{(color online) Energy-distribution curves of the (001) surface
spectral function of LaPtBi close to the $\bar{\Gamma}$ point along high-symmetry directions: (a) $\left(x00\right)$ direction ($k_y=0$), (b) $\left(xx0\right)$ direction ($k_x=k_y$), (c) $\left(x-x0\right)$ direction ($k_x=-k_y$). Weyl point projections are highlighted by grey dots. Note the Kramers doublet and the associated surface Dirac cone at $\bar{\Gamma}$. }
\label{fig:LaPtBi_BZ_cuts}
\end{figure}

The additional Fermi pocket indeed originates from a Dirac cone as can be seen in Fig.~\ref{fig:LaPtBi_BZ_cuts}, where we show energy-distribution curves (EDC) of the (001) surface
spectral function close to the $\bar{\Gamma}$ point along the $\left(x00\right)$ direction ($k_y=0$), along the $\left(xx0\right)$ direction ($k_x=k_y$) and along the $\left(x-x0\right)$ direction ($k_x=-k_y$).
As we can see, above $E=0$ there is a surface Kramers doublet pinned at $\bar{\Gamma}$ which is the vertex of a surface Dirac cone with a linear dispersion. Close to the doublet, the Fermi level cuts out a closed Fermi pocket. In particular, the states along the pocket do not contribute to the Fermi arcs connecting the Weyl nodes. However, at lower energies the presence of the bulk Weyl nodes makes it possible to ``open'' the surface cone. At this point the former Dirac states become Fermi arc states. This demonstrates the close connection between Fermi arc states and Dirac states on the surface of time-reversal invariant Weyl semimetals. 

\begin{figure}[t]\centering
\includegraphics[width=1.0\columnwidth]
{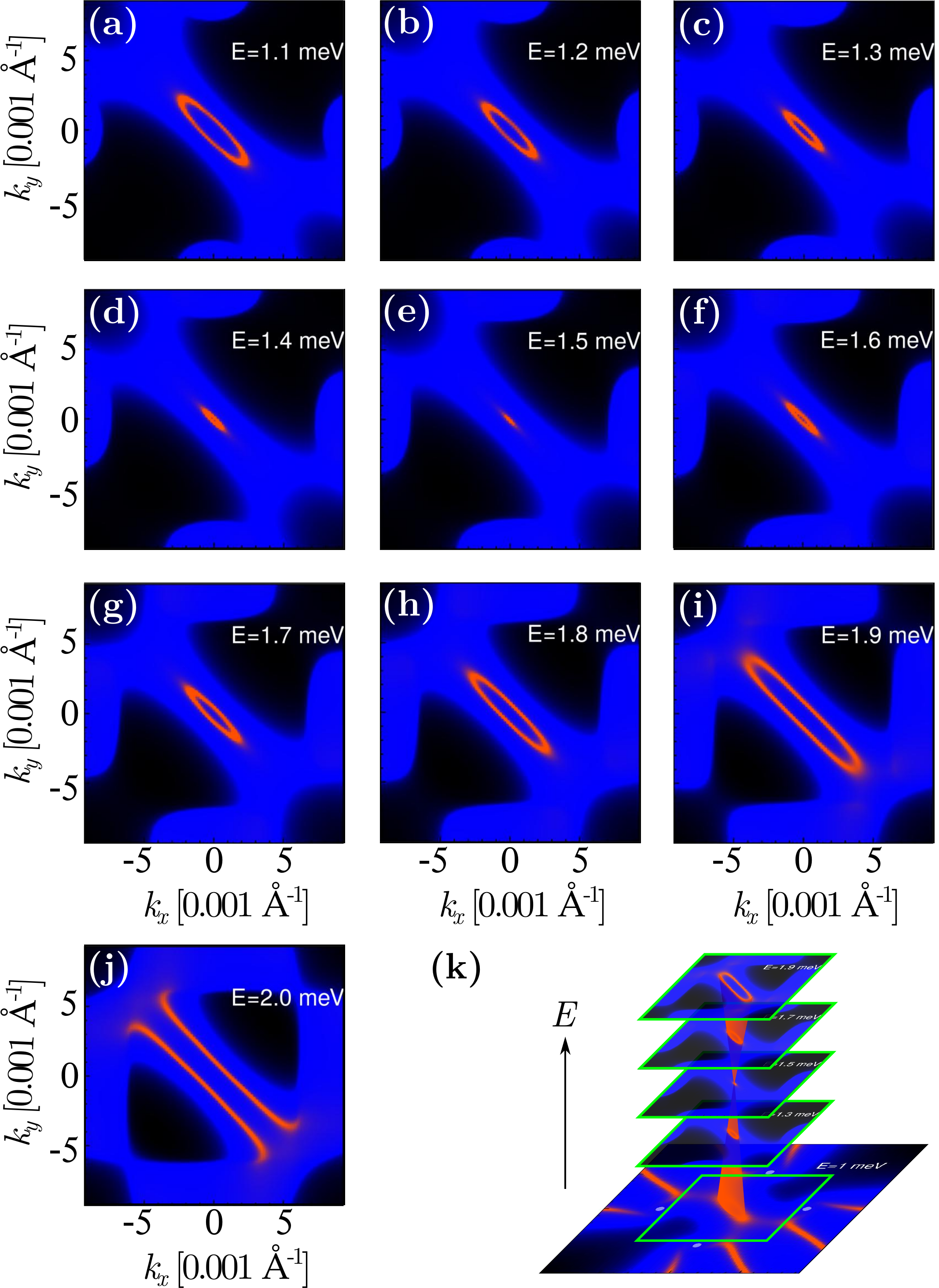}
\caption{(color online) Surface spectral weight in the (001) surface BZ of LaPtBi with varying Fermi level from (a) $E=1.1\,\mathrm{meV}$ to (j) $E=2.0\,\mathrm{meV}$ in steps of $0.1\,\mathrm{meV}$. Note how the Fermi pocket shrinks to a Dirac point in (e) and then expands again. (k) The accumulated energy slices form a Dirac cone.}
\label{fig:LaPtBi_Dirac_cone}
\end{figure}

In addition, in Fig.~\ref{fig:LaPtBi_Dirac_cone} we show a sequence of surfaces Fermi surfaces close to the $\bar{\Gamma}$ point for different Fermi levels $E_F$. We observe that the Fermi pocket shrinks as we increase the Fermi level. At $E_F\approx 1.5\,\mathrm{meV}$, the Fermi pocket has shrunk to a single point, which is the Dirac point. By tuning $E_F$ further the Fermi pocket expands again. Beyond $E_F=1.9\,\mathrm{meV}$, a second Lifshitz transition takes place: the Fermi pocket disappears, the Weyl node projections are reconnected, and the Weyl semimetal returns to its original connectivity without an additional Dirac Fermi pocket. In Fig.~\ref{fig:LaPtBi_Dirac_cone}(k), we accumulate this sequence of Fermi surfaces and indicate the surface Dirac cone.

Finally, we would like to note that our surface Fermi surfaces differ to some extent from the results shown in Ref.~\onlinecite{RJY16}. This is most likely due to differences in the details of the chosen termination.
Nonetheless, the possibility of a Fermi pocket around $\bar{\Gamma}$ is common to both studies.  

\subsection*{B: Analytical treatment of the JDOS}

In scanning-tunneling spectroscopy (STS) experiments, 
the differential conductance between the tip and the surface of a material 
is used to obtain a spatial map of the local density of electronic states (LDOS) 
at a certain energy and temperature~\cite{DML15,SVA07}. Impurities lead to characteristic modulations in the LDOS that depend strongly on the electronic structure of the host material and on the properties and distribution of impurities. 
In particular, impurities break translational symmetry on the surface thereby enabling scattering between states with the same energy $E=E_F$ but different momentum $\mathbf{k}$ and $\mathbf{k}'$. The modulations in the LDOS can be analyzed with Fourier-transform STS (FT-STS). In this way, the Fourier transformed LDOS (FTLDOS) is interpreted in terms of the quasiparticle interference (QPI) between diagonal states of the clean host material~\cite{DML15,SVA07}.

Typically, the FTLDOS is approximated by the joint density of states (JDOS)~\cite{DML15,SVA07},
which is expressed as
\begin{eqnarray}
J(\mathbf{q},E) &=& \int d^2 k\: A(\mathbf{k}+\mathbf{q},E)A(\mathbf{k},E),
\label{eq:def_JDOS}\\
A(\mathbf{k},E) &=& -1/2\pi\,\mathrm{Im}\lbrace \mathrm{Tr}[G_s(\mathbf{k},E)]\rbrace,
\end{eqnarray}
where $A(\mathbf{k},E)$ is the spectral function, $G_s(\mathbf{k},E)$ is the surface 
Green's function, and $\mathbf{k}$ is the momentum parallel to the surface.

\begin{figure}[t]\centering
\includegraphics[width=0.95\columnwidth]
{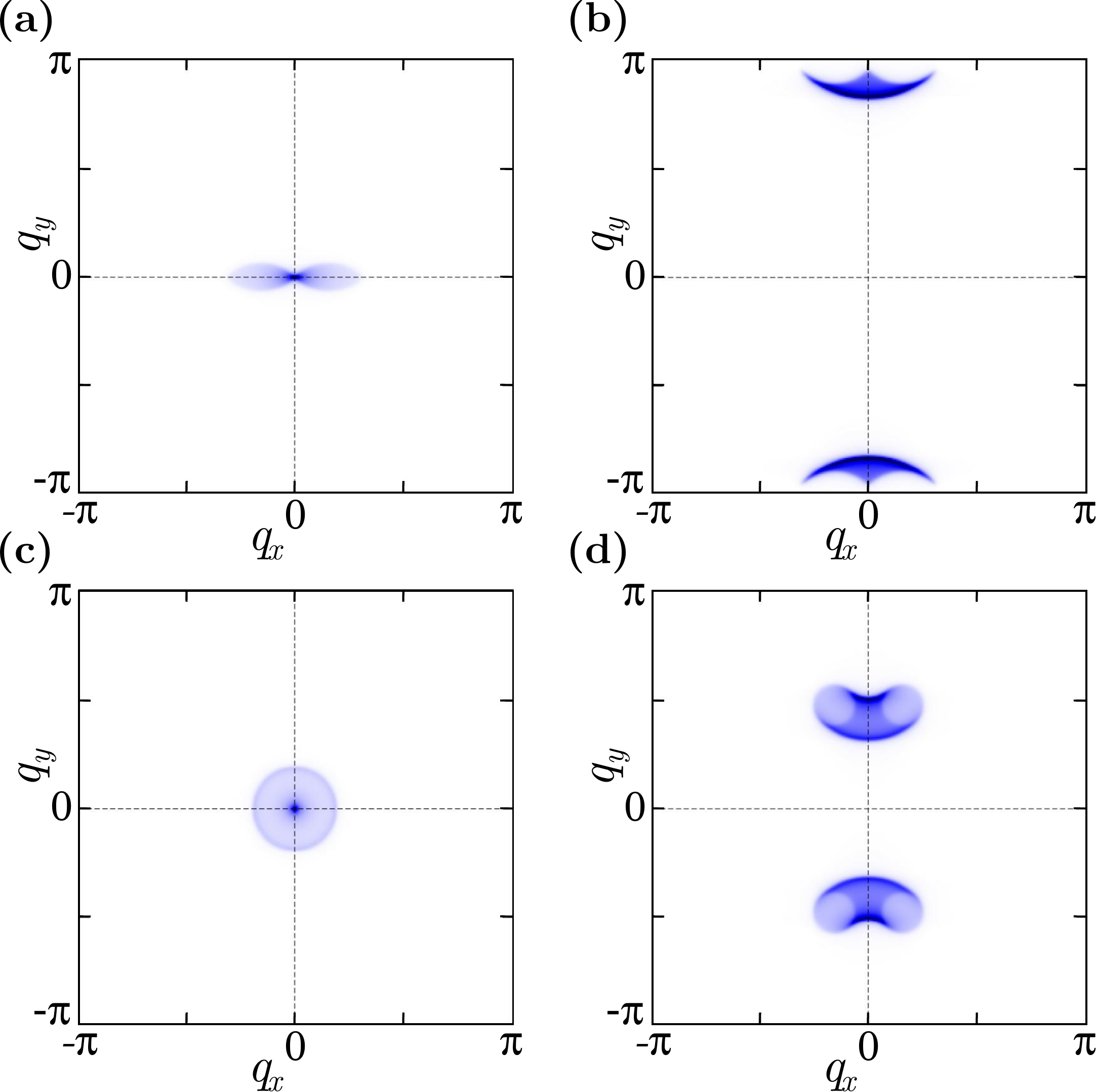}
\caption{(color online) JDOS contributions from an analytical
treatment: (a) intra-arc scattering, (b) inter-arc scattering, (c)
intra-pocket scattering, (d) arc-pocket scattering. Note that in each panel the color scale is renormalized with respect to the considered JDOS contribution. The color scale ranges from white (small JDOS), over blue to black (high JDOS).}
\label{fig:JDOS_analytical_contributions}
\end{figure} 

In order to obtain a phenomenological picture of the QPI patterns on the surface of Weyl
semimetals, we consider an idealized Fermi surface consisting of a circular Fermi pocket around the origin 
which is surrounded by an open Fermi arc of constant curvature and its time-reversal partner
[see Fig.~2(c) in the main text]. Furthermore, we assume constant spectral density.
The spectral function of the system can then be decomposed as $A=A_p+A_{a_1}+A_{a_2}$ with
\begin{eqnarray}
A_p(\mathbf{k}) &=& \int_{0}^{2\pi} d\alpha'\: \delta(\mathbf{k}-R_p\,(\cos\alpha',\sin\alpha')]),
\label{eq:A_p}\\
A_{a_{1,2}}(\mathbf{k}) &=& \int_{\alpha}^{\alpha+\Delta\alpha}\!\!\!\!\!\!\!\!\!\!
d\alpha'\: \delta(\mathbf{k}\mp[\mathbf{K}\! + \!R_a\,(\cos\alpha',\sin\alpha')]).
\label{eq:A_a}
\end{eqnarray}
$A_p$ is the spectral function associated with the circular Fermi pocket of radius $R_p$. $A_{a_{1,2}}$ are the contributions from the two Fermi arcs. The arcs are cut out from circles centered at $\pm\mathbf{K}$ with radius $R_a$. The end points of the arcs are at $\pm\alpha$ and $\pm\alpha\pm\Delta\alpha$. With this decomposition, the
JDOS of Eq.~\eqref{eq:def_JDOS} becomes
\begin{eqnarray}
J(\mathbf{q}) &=& \int d^2 k\:[
\underbrace{A_{a_1}(\mathbf{k}+\mathbf{q})A_{a_1}(\mathbf{k}) 
+ A_{a_2}(\mathbf{k}+\mathbf{q})A_{a_2}(\mathbf{k})}_\textrm{intra-arc}
\nonumber\\
{}&& + \underbrace{A_{a_1}(\mathbf{k}+\mathbf{q})A_{a_2}(\mathbf{k})
+ A_{a_2}(\mathbf{k}+\mathbf{q})A_{a_1}(\mathbf{k})}_\textrm{inter-arc}
\nonumber\\
{}&& + \underbrace{A_p(\mathbf{k}+\mathbf{q})[A_{a_1}(\mathbf{k}) + A_{a_2}(\mathbf{k})]}_\textrm{arc-pocket}
\nonumber\\
{}&& + \underbrace{[A_{a_1}(\mathbf{k}+\mathbf{q}) + A_{a_2}(\mathbf{k}+\mathbf{q})]A_p(\mathbf{k})}_\textrm{arc-pocket}
\nonumber\\
{}&& + \underbrace{A_p(\mathbf{k}+\mathbf{q})A_p(\mathbf{k})}_\textrm{intra-pocket}].
\end{eqnarray}
We see that there are four different contributions to the total JDOS which can be attributed to scattering events within each Fermi arc (intra-arc), between the Fermi arcs (inter-arc), between Fermi pocket and Fermi arcs (arc-pocket), and within the Fermi pocket (intra-pocket). Inserting Eqs.~\eqref{eq:A_p} and~\eqref{eq:A_a}, each individual contribution involves integrals of the form
\begin{equation}
\int d^2 k \int d\alpha'\, d\alpha''\: \delta[\mathbf{k}+\mathbf{q}-\mathbf{u}_i(\alpha')]\,
\delta[\mathbf{k}-\mathbf{u}_j(\alpha'')].
\end{equation}
This can be written as a convolution of two $\delta$ functions, $\delta\ast\delta$. By using that 
$\delta\ast\delta\equiv\delta$, the integral becomes
\begin{equation}
\int d\alpha'\, d\alpha''\: \delta[\mathbf{q}+\mathbf{u}_i(\alpha')]\,
\delta[\mathbf{u}_j(\alpha'')].
\end{equation}
In general, this integral cannot be simplified further. To obtain qualitative results for the JDOS,
we therefore approximate the appearing $\delta$ functions of the form $\delta(\mathbf{k})\equiv\delta(k_x)\delta(k_y)$
by Lorentz functions
\begin{equation}
\delta_\epsilon(k_i) = \frac{1}{\pi}\,\frac{\epsilon}{k_i^2 + \epsilon^2},\:\:\:\: \epsilon\ll 1,
\end{equation}
with height $1/\sqrt{\epsilon}$ and width $\sqrt{\epsilon}$. The ensuing integrals are then solved 
numerically. The results for the different JDOS contributions are shown in Fig.~\ref{fig:JDOS_analytical_contributions}. Note that in each panel the color scale is renormalized 
with respect to the considered contribution.

\begin{figure}[t]\centering
\includegraphics[width=1.0\columnwidth]
{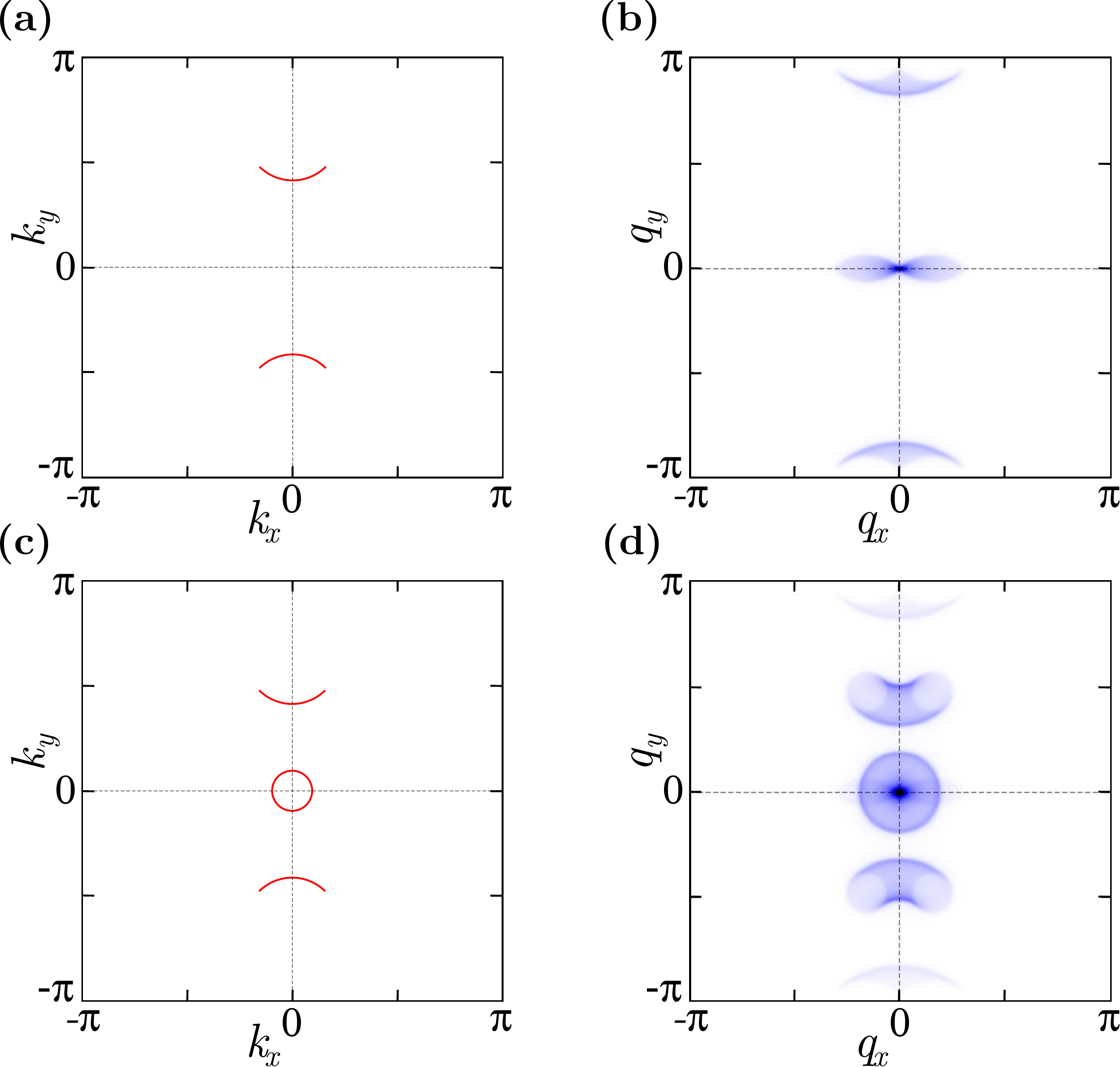}
\caption{(color online) JDOS contributions from an analytical
treatment: (a) Fermi surface consisting of two Fermi arcs only. The corresponding JDOS in (b) shows the characterisitc pinch point at $\mathbf{q}=0$. (c) A Fermi surface with additional circular Fermi pocket shows additional features in the JDOS as shown in (d). Most remarkable are the kidney-shaped features stemming from scattering between Fermi pocket and Fermi arcs.}
\label{fig:JDOS_analytical_total}
\end{figure} 

Fig.~\ref{fig:JDOS_analytical_contributions}(a) shows the intra-arc JDOS contributions. We can clearly see 
a pinch point at $\mathbf{q}=0$ which is in the center of a dumbell-shaped feature aligned with the $q_x$ axis. This is the unique QPI pattern of an open Fermi arc. Since the considered
Fermi surface consists of two of those arcs, we get an additional JDOS (inter-arc) contribution originating from scattering between the arcs. These features appear around the $q_y$ axis at $\mathbf{q}$ vectors corresponding to scattering vectors between the two arcs in the BZ [see Fig.~\ref{fig:JDOS_analytical_contributions}(b)].

The intra-cone JDOS contribution, shown in Fig.~\ref{fig:JDOS_analytical_contributions}(c), is strongly peaked at $\mathbf{q}=0$ and falls off rapidly away from the center. The resulting pattern is rotationally symmetric reflecting the symmetry of the Fermi pocket. The feature has a sharp boundary with a slightly enhanced intensity
at $|\mathbf{q}|$ values corresponding to the diameter of the underlying Fermi pocket. This is due to scattering events from opposing states of the Fermi pocket.

In Fig.~\ref{fig:JDOS_analytical_contributions}(d) we present the arc-pocket JDOS contribution which is characteristic of the coexistence of open Fermi arcs and closed Fermi pockets. It therefore represents the universal QPI pattern of Weyl semimetals with additional surface Dirac cones. The kidney-shaped features appear
around the $q_y$ axis at $\mathbf{q}$ vectors corresponding to scattering vectors that connect the open Fermi arcs to the closed Fermi pocket. 

All separate JDOS contributions presented in Figs.~\ref{fig:JDOS_analytical_contributions}(a)-(d) have very different intensities. This can be accounted for by the phase-space volume available for scattering events contributing to the considered JDOS contribution. Accordingly, the intra-pocket features are very high in intensity whereas the inter-arc contributions are very low. This is most apparent in the accumulated JDOS spectra which we show in Figs.~\ref{fig:JDOS_analytical_total}(b) and~(d). 

In Fig.~\ref{fig:JDOS_analytical_total}(b), we show the accumulated JDOS taking into account only intra- and inter-arc contributions, which corresponds to a Fermi surface consisting only of two open Fermi arcs as depicted in Fig.~\ref{fig:JDOS_analytical_total}(a). On the contrary, in Fig.~\ref{fig:JDOS_analytical_total}(d) we accumulate all four JDOS contributions associated with a Fermi surface comprising two Fermi arcs and an additional Fermi pocket in the center [see Fig.~\ref{fig:JDOS_analytical_total}(c)]. We can clearly see how the total JDOS is dominated by QPI features stemming from scattering events involving the Fermi pocket.

\subsection*{C: Wannier-center formulation of the $\Z_2$ invariant}

In the following, we are going to review a method, introduced in Ref.~\onlinecite{YQB11}, that enables us to calculate the topological $\Z_2$
invariant of a general 2D insulator with time-reversal symmetry. This method allows us to calculate the $\Z_2$ number without choosing a gauge-fixing condition, which makes it particularly appealing for numerical studies.

The central notion of this method is the \emph{time-reversal polarization}~\cite{FuK06}. In a time-reversal invariant 1D band insulator, the $2N$ occupied bands can be decomposed into two sets which are connected by time reversal. The time-reversal polarization is then the difference in the net charge polarization of the two sets of bands. Furthermore, it can be shown that the time-reversal polarization can only assume the values 0 or 1 (modulo 2) as long as the system preserves time-reversal symmetry.

A 2D time-reversal invariant (TRI) band insulator can be thought of as a collection of 1D insulators in momentum space parametrized by, say, $k_y$. In this collection, only the effective 1D systems at $k_y=0$ and $\pi$ preserve 1D time-reversal symmetry. Thus, the time-reversal polarization can assume non-integer values in between. The key idea is now to determine how the time-reversal polarization changes with $k_y$. It was shown that this leads to a well-defined $\Z_2$ invariant for a 2D band insulator~\cite{YQB11,FuK06}. Moreover, the charge polarization of an occupied band is related to its Wannier center in position space. For this reason, the change in time-reversal polarization can also be understood as a shift of the Wannier centers.

The Wannier centers of the occupied bands at fixed $k_y$ are defined as the eigenvalues of the position operator projected onto the occupied subspace. By using localized Wannier functions $|\alpha j\rangle$ as a basis, where $\alpha$ is an orbital index and $j$ denotes the lattice site, the position operator for a 1D lattice system with periodic boundary conditions can be written as
\begin{equation}
\hat{X} = \sum_{\alpha j} e^{-i\frac{2\pi j}{N_x}}\,|\alpha j\rangle \langle\alpha j|,
\end{equation}
where $N_x$ is the number of unit cells. Furthermore, the projection operator onto the occupied bands $o$ at fixed $k_y$ is defined as
\begin{equation}
\hat{P}(k_y) = \sum_{n\in o,k_x} |\Psi_{n k_x k_y}\rangle \langle \Psi_{n k_x k_y}|,
\end{equation}
with the Bloch state $|\Psi_{n \mathbf{k}}\rangle = e^{i\mathbf{k}\cdot\mathbf{r}} |n \mathbf{k}\rangle$.
The projected position operator can be written as follows
\begin{multline}
\hat{X}_P(k_y) = \hat{P}(k_y)\hat{X}\hat{P}(k_y)\\
= \sum_{j=1}^{N_x}\sum_{mn\in o}|\Psi_{n,k_{x,j},k_y}\rangle \langle \Psi_{n,k_{x,j+1},k_y}|
F^{nm}_{j,j+1}(k_y),
\end{multline}
with $k_{x,j}=2\pi j/N_x$ being the discrete $k_x$ points taken along the $x$ axis.
The $F_{j,j+1}(k_y)$ are $2N\times 2N$ matrices with matrix elements
\begin{equation}
F^{nm}_{j,j+1}(k_y) = \langle n,k_{x,j},k_y|m,k_{x,j+1},k_y \rangle.
\end{equation}
The eigenvalues of $\hat{X}_P(k_y)$ can be obtained by the transfer matrix method. For this purpose,
we define the $2N\times 2N$ matrix
\begin{equation}
T(k_y) = F_{1,2} F_{2,3} \cdots F_{N_x-1,N_x} F_{N_x,1}.
\end{equation} 
$T$ is unitary and has the following eigenvalues
\begin{equation}
\lambda_m(k_y) = e^{i\theta_m(k_y)},\:\:\: m = 1,2,\ldots,2N,
\end{equation}
which are gauge invariant under $U(2N)$ transformations of the $|n \mathbf{k}\rangle$. Furthermore,
it can be shown that in the continuum limit $T(k_y)$ corresponds to the $U(2N)$ Wilson loop~\cite{YQB11},
\begin{equation}
T(k_y) = P\,e^{-i\int_{\mathcal{C}_{k_y}}A(k_x)\,dk_x},
\end{equation}
with the non-Abelian Berry connection $A(k)$ which is a unitary $2N\times 2N$ matrix.
The eigenvalues of the projected position operator $\hat{X}_P(k_y)$, and thus the Wannier centers
of the occupied bands, are then obtained from the eigenvalues of $T$ as
\begin{equation}
\chi_{m,j}(k_y) = e^{i[\theta_m(k_y)+2\pi j]/N_x}, \:\:\: j=1,\ldots N_x.
\end{equation}
Since the Wannier centers of adjacent unit cells differ only by a constant, $k_y$ independent
phase shift $e^{i 2\pi/N_x}$, it is sufficient to look at the evolution of the $2N$ phases
$\theta_m(k_y) = \Im \log \lambda_m(k_y)$.

The connection to the $\Z_2$ invariant is established as follows. We plot the $2N$ phases $\theta_m$
as a function of $k_y$ and glue the lines $\theta=-\pi$ and $\theta=\pi$ together, such that the Wannier centers live on the surface of a cylinder. At $k_y=0$ the phases have to appear as degenerate pairs due to time-reversal symmetry. By moving away from this point, the pairs split and recombine at $k_y=\pi$ (again due to time-reversal
symmetry). Because the $\theta_m$ are phases, the Wannier center pairs may now differ by an integer multiple of $2\pi$. Hence, the evolution of the Wannier center pairs from $k_y=0$ to $k_y=\pi$ will encircle the cylinder an integer number of times. The sum of these integers over all pairs of Wannier centers defines a winding number. However, an even number of windings can always be removed, whereas this is not possible for a single winding. Consequently, the total winding number modulo 2 is a topological $\Z_2$ invariant. In practice, one draws an arbitrary reference line parallel to the $k_y$ axis and counts how many times the Wannier centers cross this line. The system is topological (trivial), if the reference line is crossed an odd (even) number of times.

\begin{figure}[t]\centering
\includegraphics[width=0.9\columnwidth]
{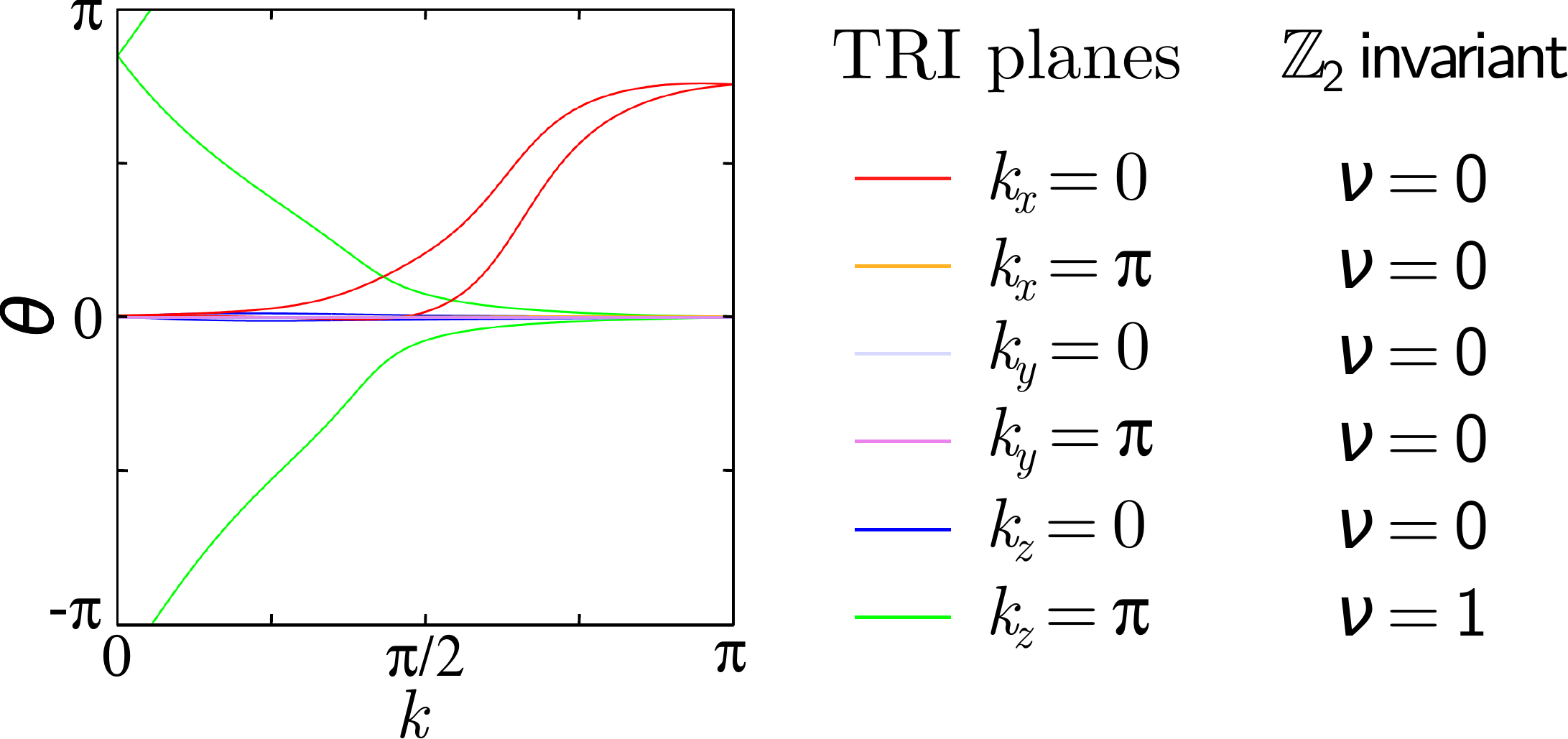}
\caption{(color online) Evolution of the Wannier centers $\theta_m$ as a function of $k$ for the TRI planes of the tight-binding model with $a=b=1$, $t=1.5$, $\alpha=0.3$, $d=0.1$, $\lambda=0.5$ and $\beta=0.7$. $k$ denotes a momentum parallel to the considered plane in the 3D BZ. We also indicate the corresponding value of the $\Z_2$ invariant $\nu$ inferred from the winding number of the Wannier pair.}
\label{fig:Wannier_centers}
\end{figure} 
As an example, in Fig.~\ref{fig:Wannier_centers} we show the evolution of the $\theta_m$ for the six TRI planes of the tight-binding model considered in the main text. There are only two occupied bands which is why we only need to look at a single pair of Wannier centers for each plane. We find that only the Wannier pair for the TRI plane at $k_z=\pi$ winds non-trivially around the $\theta$-$k$ cylinder. Hence, we infer that the $\Z_2$ invariant of this plane is $\nu=1$. All other TRI planes have a trivial $\Z_2$ number, i.e., $\nu=0$. 

\begin{figure}[t]\centering
\includegraphics[width=0.9\columnwidth]
{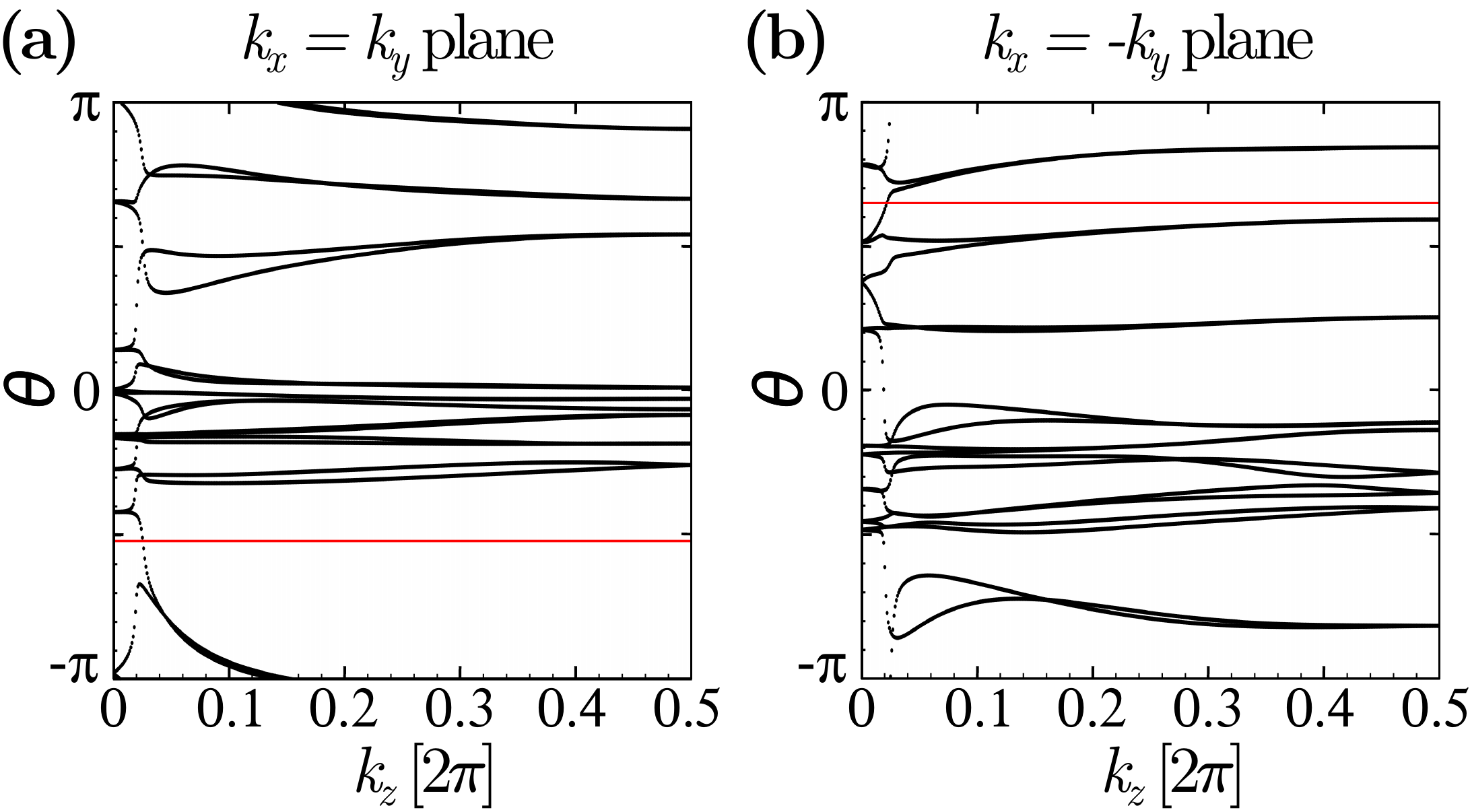}
\caption{(color online) Evolution of the Wannier centers $\theta$ as a function of $k_z$ for the time-reversal invariant planes (a) $k_x=k_y$, and (b) $k_x=-k_y$ of LaPtBi. The path along $k_z$ ranges from $\Gamma$ ($k_z=0$) to $\Z$ ($k_z=\pi$) of the bulk BZ. A reference line (red) is used to determine the $\Z_2$ topological invariant of the planes.}
\label{fig:Wannier_centers_LaPtBi}
\end{figure}

Furthermore, we have used this scheme to determine the $\Z_2$ invariants of the $k_x=\pm k_y$ planes in our DFT study of LaPtBi. The evolution of the Wannier centers is shown in Fig.~\ref{fig:Wannier_centers_LaPtBi}. In both cases, the chosen reference line is crossed only once. From this we infer that both time-reversal invariant planes are topological nontrivial with $\Z_2$ invariants $\nu=1$.

\subsection*{D: Iterative scheme for the calculation of the surface Green's function}

In order to calculate the surface JDOS of a lattice system, we need to compute its corresponding surface Green's function. In the following, we are going to review an iterative scheme, introduced in Ref.~\onlinecite{SSR85}, for the computation of this surface Green's function. For simplicity, we will restrict our considerations to systems described by tight-binding Hamiltonians with only nearest-neighbor hopping terms. We note, however, that any solid with a surface can be reduced to a semi-infinite stack of principal layers with nearest-neighbor interactions~\cite{SSR85}.

Let us consider a tight-binding model described by the lattice Hamiltonian $\mathcal{H}$. A surface can be introduced by cleaving the lattice along a principal layer of unit cells. Each unit cell in this layer may have $M$ degrees of freedom, such as spin, orbital or sublattice degrees of freedom. Let us further assume that the system preserves translational symmetry parallel to the surface. Consequently, the momentum $\mathbf{k}_{||}$ parallel to the surface is a good quantum number and we can form Bloch-state vectors for each principal layer of the form
\begin{equation}
\Psi_n(\mathbf{k}_{||}) = (\varphi_n^1(\mathbf{k}_{||}),\ldots,\varphi_n^M(\mathbf{k}_{||})),
\end{equation}
where $n$ labels the layer. By taking matrix elements of the operator equation
$(\omega-\mathcal{H})\mathcal{G}(\omega)=\mathcal{I}$ with the Bloch states, we get the following chain of coupled equations for each $\mathbf{k}_{||}$
\begin{eqnarray}
(\omega-H_{00})G_{00} &=& \id + H_{01}G_{10}\label{eq:system_G00}\\
(\omega-H_{00})G_{10} &=& H_{01}^\dagger G_{00} + H_{01}G_{20}\\
&\vdots &\nonumber\\
(\omega-H_{00})G_{n0} &=& H_{01}^\dagger G_{n-1,0} + H_{01}G_{n+1,0},
\label{eq:system_Gn0}
\end{eqnarray}
where the $H_{nm}$ and $G_{nm}$ are $M\times M$ matrices defined as
\begin{eqnarray}
H_{nm}(\mathbf{k}_{||}) &=& \langle\Psi_n(\mathbf{k}_{||})|\mathcal{H}|\Psi_m(\mathbf{k}_{||})\rangle,\\
G_{nm}(\omega,\mathbf{k}_{||}) &=& \langle\Psi_n(\mathbf{k}_{||})|\mathcal{G}(\omega)|\Psi_m(\mathbf{k}_{||})\rangle,
\end{eqnarray}
and we have assumed an ideal surface with $H_{00}=H_{11}=\ldots=H_{nn}$ and $H_{01}=H_{12}=\ldots=H_{n-1,n}$. Note that $n=0$ corresponds to the surface principal layer. Hence, $G_{00}(\omega,\mathbf{k}_{||})$ defines the surface Green's function.

The general equation for $G_{n0}$ in Eq.~\eqref{eq:system_Gn0} can be rewritten as ($n\geq 1$)
\begin{equation}
G_{n0}(\omega) = (\omega-H_{00})^{-1}(H_{01}^\dagger G_{n-1,0} + H_{01} G_{n+1,0}).
\label{eq:general_Gn0}
\end{equation}
If we put $n=1$ in this equation and plug it into Eq.~\eqref{eq:system_G00} we get
\begin{multline}
[\omega - H_{00} - H_{01}(\omega-H_{00})^{-1} H_{01}^\dagger] G_{00}\\
 = \id + H_{01}(\omega-H_{00})^{-1}H_{01}G_{20}.
\end{multline}
Similarly, we can replace $G_{n-1,0}$ and $G_{n+1,0}$ in Eq.~\eqref{eq:system_Gn0} by replacing
$n\rightarrow n-1$ and $n\rightarrow n+1$, respectively. The ensuing equations can be written
more compactly as ($n\geq 2$)
\begin{eqnarray}
(\omega - \epsilon_{1s})G_{00} &=& \id + \alpha_1 G_{20} \\
(\omega - \epsilon_{1})G_{n0} &=& \beta_1 G_{n-2,0} + \alpha_1 G_{n+2,0}
\end{eqnarray}
with 
\begin{eqnarray}
\alpha_1 &=& H_{01}(\omega-H_{00})^{-1} H_{01}\\
\beta_1 &=& H_{01}^\dagger(\omega-H_{00})^{-1} H_{01}^\dagger\\
\epsilon_{1s} &=& H_{00} + H_{01}(\omega-H_{00})^{-1} H_{01}^\dagger\\
\epsilon_1 &=& H_{00} + H_{01}(\omega-H_{00})^{-1} H_{01}^\dagger \nonumber\\
&&{} + H_{01}^\dagger(\omega-H_{00})^{-1} H_{01}.
\end{eqnarray}
These equations only involve next-nearest-neighbor
Green's functions, whereas nearest neighbors have disappeared completely.

Let us now consider the subset of equations formed by taking only even values for $n$, namely
\begin{eqnarray}
(\omega - \epsilon_{1s})G_{00} &=& \id + \alpha_1 G_{20} \\
(\omega - \epsilon_{1})G_{20} &=& \beta_1 G_{00} + \alpha_1 G_{40} \\
&\vdots &\nonumber\\
(\omega - \epsilon_{1})G_{2n,0} &=& \beta_1 G_{2(n-1),0} + \alpha_1 G_{2(n+1),0}.
\end{eqnarray}
This set of equations is isomorphic to Eqs.~\eqref{eq:system_G00}--\eqref{eq:system_Gn0} except for
the different zeroth-order matrix elements, $\epsilon_{1s}\neq\epsilon_1$. Thus, we can reiterate the same steps to obtain $\alpha_2$, $\beta_2$, $\epsilon_2$ and $\epsilon_{2s}$. Starting with $\epsilon_0=\epsilon_{0s}=H_{00}$, $\alpha_0=H_{01}$ and $\beta_0=H_{01}^\dagger$, this defines an iterative sequence in which after the $i$-th step the system of equations reads
\begin{eqnarray}
(\omega - \epsilon_{is})G_{00} &=& \id + \alpha_i G_{2^i,0} \label{eq:G_00_iterated}\\
(\omega - \epsilon_{i})G_{2^i,0} &=& \beta_i G_{00} + \alpha_1 G_{2^{i+1},0} \\
&\vdots &\nonumber\\
(\omega - \epsilon_{i})G_{2^in,0} &=& \beta_i G_{2^i(n-1),0} + \alpha_i G_{2^i(n+1),0}
\end{eqnarray}
with 
\begin{eqnarray}
\alpha_i &=& \alpha_{i-1}(\omega-\epsilon_{i-1})^{-1} \alpha_{i-1}\\
\beta_i &=& \beta_{i-1}(\omega-\epsilon_{i-1})^{-1} \beta_{i-1}\\
\epsilon_i &=& \epsilon_{i-1} + \alpha_{i-1}(\omega-\epsilon_{i-1})^{-1} \beta_{i-1}\nonumber\\
&&{}+ \beta_{i-1}(\omega-\epsilon_{i-1})^{-1} \alpha_{i-1}\\
\epsilon_{is} &=& \epsilon_{i-1,s} + \alpha_{i-1}(\omega-\epsilon_{i-1})^{-1} \beta_{i-1}.
\end{eqnarray}
After the $i$-th step, this set of equations describes an effective Hamiltonian for a chain with a unit cell enlarged by a factor of $2^i$, with nearest-neighbor interactions $\alpha_i$ and $\beta_i$, and with 
zeroth-order Hamiltonian matrix elements $\epsilon_i$ and $\epsilon_{is}$. In each layer of this effective chain
the effects of nearest-neighbor interactions of all the previous chains are implicitely encoded. Therefore, the norm of the $\alpha$'s and $\beta$'s typically decreases with $i$. Once they are small enough, we have $\epsilon_i\simeq\epsilon_{i-1}$ and $\epsilon_{is}\simeq\epsilon_{i-1,s}$. In particular, the RHS of Eq.~\eqref{eq:G_00_iterated} becomes $\simeq\id$ and we can finally solve for the surface Green's function,
\begin{equation}
G_S \equiv G_{00}(\omega) \simeq (\omega - \epsilon_{is})^{-1}.
\end{equation}
Thus, we have obtained a good approximation for $G_{00}$. We remark that the Green's function of the dual (or opposite) surface can be obtained be exchanging the role of $\alpha_i$ and $\beta_i$. The Green's function of the bulk is obtained in this process as
\begin{equation}
G_b\equiv \lim_{n\rightarrow\infty}G_{nn} \simeq G_{2^i,2^i} \simeq (\omega-\epsilon_i)^{-1}.
\end{equation}
In practice, one usually wishes to determine
the retarded or advanced Green's function. For that, we simply replace in the iterative scheme $\omega$ by $\omega\mp i\eta$, with $\eta\in\R$ sufficiently small. 
Moreover, we note that the great advantage of this scheme is its fast convergence. Typically, one achieves $\beta_i\simeq 0$, $\alpha_i \simeq 0$ to numerical accuracy in fewer than 10 iterations.  

\subsection*{E: Surface energy spectra of the tight-binding model}

\begin{figure}[t]\centering
\includegraphics[width=1.0\columnwidth]
{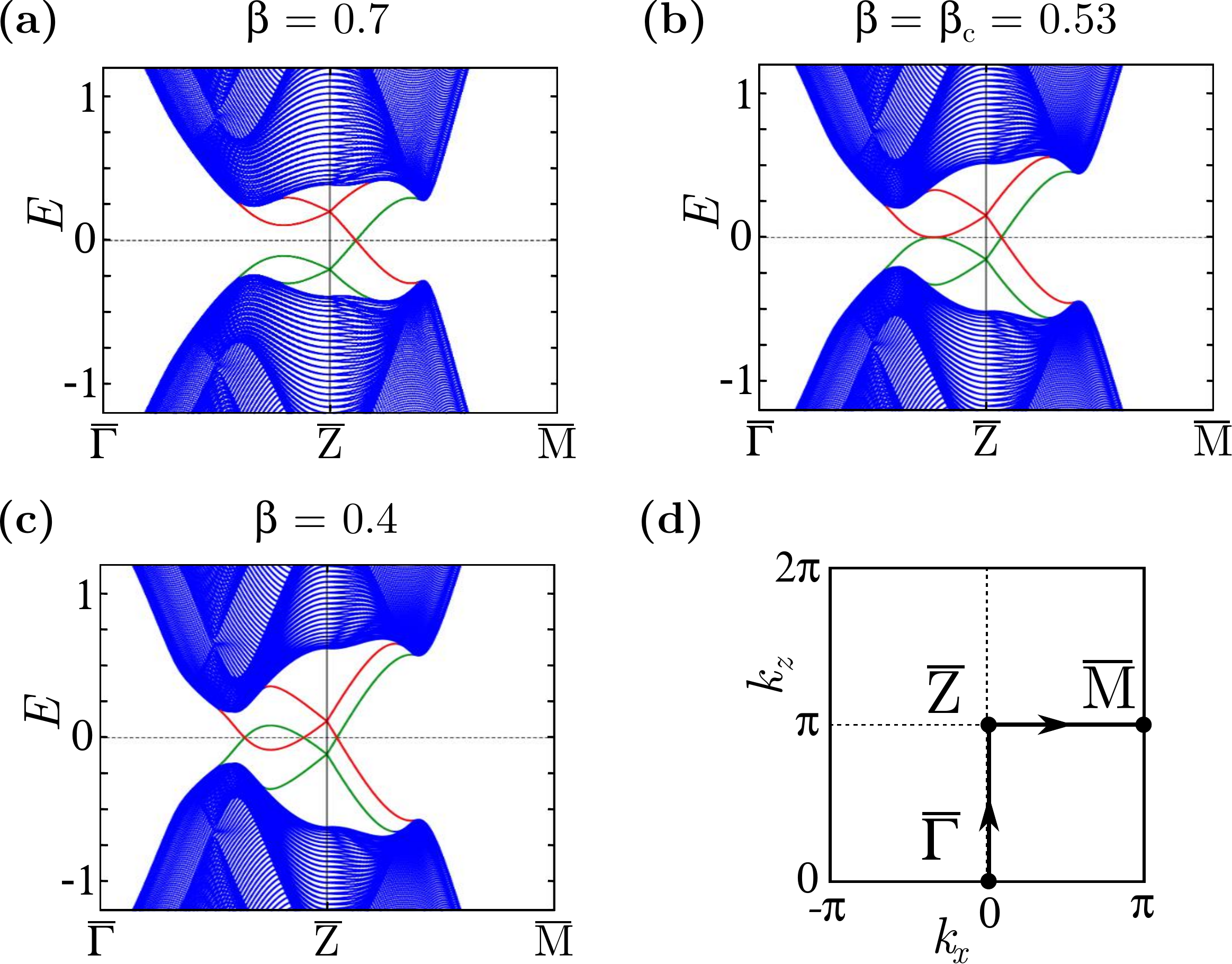}
\caption{(color online) Surface energy spectra along high-symmetry lines in the tight-binding model with $a=b=1$, $t=1.5$, $\alpha=0.3$, $d=0.1$, $\lambda=0.5$: (a)--(c) energy spectra for different values of the parameter $\beta$. Bulk-like states are highlighted in blue, surface states are highlighted in red and green indicating different surfaces. The Fermi energy is at $E_F=0$ (dashed line). (d) High-symmetry path through the surface BZ corresponding to the displayed energy spectra.}
\label{fig:tm_model_bandstructures}
\end{figure}

In Fig.~\ref{fig:tm_model_bandstructures}, we present energy spectra of the tight-binding model studied in the main text. In particular, Figs.~\ref{fig:tm_model_bandstructures}(a)--(c) show high-symmetry lines through the $k_x$-$k_z$ surface BZ which are indicated in Fig.~\ref{fig:tm_model_bandstructures}(d).

Recall that for $\beta=0.7$ the surface Fermi surface consists of two Fermi arcs only. They intersect only the $k_z=\pi$ axis in two points that are related by time reversal. The corresponding energy spectrum along $\bar{\Gamma}\bar{Z}$ and $\bar{Z}\bar{M}$ is shown in Fig.~\ref{fig:tm_model_bandstructures}(a). For the sake of clarity, let us focus on surface states associated with only one of the two surfaces. We see that at $\bar{Z}$ there is an in-gap surface Kramers doublet, relatively far away from the Fermi energy, which is protected by time-reversal symmetry. Away from this point, the doublet is split but the ensuing non-degenerate states evolve differently depending on the considered direction in $\mathbf{k}$ space. Along $\bar{\Gamma}\bar{Z}$ both surface bands terminate at the bulk conduction/valence band. On the contrary, along $\bar{Z}\bar{M}$ one band terminates at the conduction band whereas the other terminates at the valence band. By bulk-boundary correspondence, this behavior is in agreement with the topological invariants of the 2D bulk systems associated with these lines.

By decreasing $\beta$, the Kramers doublet at $\bar{Z}$ moves closer to the Fermi energy. At the Lifshitz transition [see Fig.~\ref{fig:tm_model_bandstructures}(b)], one of the surface bands along $\bar{\Gamma}\bar{Z}$ intersects the Fermi-energy plane tangentially. This is the point where the topology of the Fermi surfaces changes: 
below a crictical $\beta_c$, the Kramers doublet, whose low-energy spectrum is a 2D Dirac cone, is close enough to the Fermi-energy plane to cut out a circular Fermi pocket [see Fig.~\ref{fig:tm_model_bandstructures}(c)]. To preserve the evenness of the number of crossings, there also has to be a second crossing along $\bar{\Gamma}\bar{Z}$. In contrast, the number of $E=0$ states along $\bar{Z}\bar{M}$ has not changed. This leads to the formation of new Fermi arcs which intersect only the $k_x=0$ line of the surface BZ.

Furthermore, we note that the Dirac cone appearing on the surface of the considered Weyl semimetal has a peculiar feature: along one direction, here $\bar{Z}\bar{M}$, it connects the bulk valence to the bulk conduction band, as on the surface of a topological insulator. However, along the perpendicular direction, here $\bar{\Gamma}\bar{Z}$, the Dirac cone is connected to the bulk conduction band (or to the bulk valence band) only. Such a ``dangling" Dirac cone can only appear on the surface of a Weyl semimetal where the existence of additional bulk Weyl cones and surface Fermi arcs, to which the Dirac cone is continuously connected in the 2D surface BZ, resolves this apparent contradiction.

\begin{figure}[t]\centering
\includegraphics[width=0.95\columnwidth]
{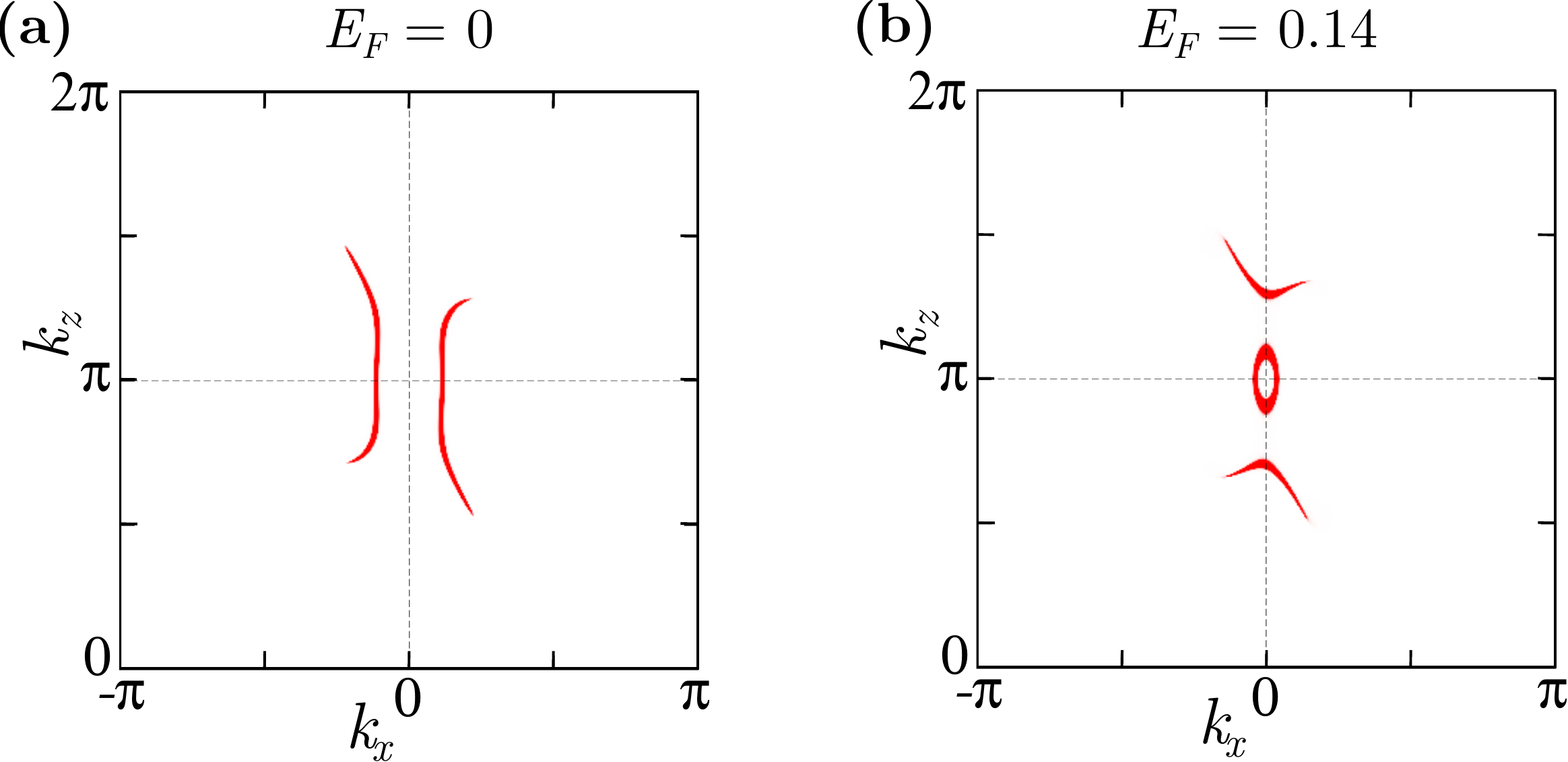}
\caption{(color online) Surface spectral weight for the (010) surface in the tight-binding model with $a=b=1$, $t=1.5$, $\alpha=0.3$, $d=0.1$, $\lambda=0.5$ and $\beta=0.7$. Note that the Fermi energy $E_F$ is varied while the parameter $\beta$ is fixed. We can see how the Fermi arc connectivity is changed by varying the Fermi energy.}
\label{fig:tm_model_Fermi_surface_EF}
\end{figure}

Finally, we show that the Fermi arc connectivity can also be changed by varying the Fermi energy $E_F$. For that, we analyze the surface spectral weight for the (010) surface of the tight-binding model (see Fig.~\ref{fig:tm_model_Fermi_surface_EF}). Since only the Fermi energy is varied, the progression can be compared to the energy dispersion of the red surface bands in Fig.~\ref{fig:tm_model_bandstructures}(a). For $E_F=0$, the Dirac point is far above the Fermi energy and there are no surface states at the Fermi energy along $\bar{\Gamma}\bar{Z}$. Hence, there is no Dirac Fermi pocket as can be seen in Fig.~\ref{fig:tm_model_Fermi_surface_EF}(a). By increasing $E_F$, the Fermi level moves closer to the Dirac point until it intersects the ``dangling'' arm of the Dirac cone along $\bar{\Gamma}\bar{Z}$. This is where the Lifshitz transition takes place and the connectivity of the Fermi arcs is changed in the same way as for varying only the paramter $\beta$. In particular, a Dirac Fermi pocket is formed [see Fig.~\ref{fig:tm_model_Fermi_surface_EF}(b)]. The linear dispersion of the corresponding Dirac states can be inferred from Fig.~\ref{fig:tm_model_bandstructures}(a).

The process of inducing the Lifshitz transition by varying the Fermi energy $E_F$ is equivalent to the one discussed in the main part of the Letter, where we varied the parameter $\beta$ instead. Furthermore, this shows qualitatively how the Lifshitz transition in the candidate material LaPtBi takes place: LaPtBi already posesses a Dirac point away from $E_F=0$, as can be seen in Fig.~\ref{fig:LaPtBi_BZ_cuts}. By shifting the Fermi level towards positive energies, we enter the interval around the Dirac point in which the states form a closed cone and thus, a Dirac Fermi pocket appears in the surface BZ.

\subsection*{F: Independent $\Z_2$ invariants for TRI Weyl semimetals}

As opposed to TRI insulators in 3D, the $\Z_2$ invariants in TRI Weyl semimetals are in general independent. The reason is the breakdown of the 3D homotopy argument\cite{MoB07,Roy09} due to the gapless nature of Weyl semimetals. 

The six TRI planes as defined in Refs.~\onlinecite{MoB07,Roy09} can be considered as independent 2D systems because they intersect mutually along only a single line at most. In general, a generic gap closing in one plane, which changes the 2D $\Z_2$ invariant, can therefore not affect any of the other planes. 

However, if the 3D system accomodating the six TRI planes has a full bulk energy gap, a 3D $\Z_2$ invariant can be defined based on homotopy arguments. This invariant is known as the ``strong'' invariant of TRI insulators. The value of this \emph{one} invariant determines how the \emph{six} 2D $\Z_2$ invariants of the TRI planes are related~\cite{MoB07,Roy09}. In particular, it gives rise to \emph{three} additional constraints which reduces the total number of independent invariants to $(6+1)-3=4$ invariants. Typically, these invariants are taken to be the strong 3D invariant $\nu_0$ and the three ``weak'' 2D invariants $(\nu_1\nu_2\nu_3)$ associated with the TRI planes forming the borders of the BZ~\cite{FKM07}. The remaining $\Z_2$ invariants of the other TRI planes can then be calculated only from the knowledge of these four numbers.

On the contrary, this is not possible for a time-reversal invariant WSM. First of all, the strong 3D invariant is not defined simply because a WSM is gapless by definition. A homotopy argument would require the existence of a full energy gap. As a consequence, there are no longer additional constraints on the values of the six 2D $\Z_2$ invariants. Hence, the six $\Z_2$ invariants for the TRI planes are in general independent for time-reversal invariant WSMs.

\subsection*{G: Connection between Fermi pocket and Dirac cone}

In the following, we are going to discuss the relation between a Fermi pocket and the presence of a Dirac cone in more detail. 
 
Due to time-reversal symmetry, there are 8 TRI momenta in the bulk BZ and 4 TRI momenta in any surface BZ of a material. At these points, states must come in Kramers doublets due to the Kramers theorem. In particular, any surface state at a TRI momentum is part of a Kramers doublet. 

If there is a small Fermi pocket around a TRI point, its corresponding dispersion has to terminate in a Kramers-doublet at the TRI point if we go up or down in energy. A trivial parabolic electron or hole pocket without a band crossing is not possible, since it could be easily removed by shifting the band up or down. However, in this way the number of Kramers pairs along the projection of a corresponding TRI plane would change by \emph{one} which is forbidden by the topology of the TRI planes. The number of Kramers pairs can only change by an even number due to the $\Z_2$ nature of the 2D systems. For these reasons, there must be a Kramers-doublet band crossing associated with the Fermi pocket. 

In the vicinity of the Kramers band crossing, the Hamiltonian can be expanded with respect to $\mathbf{k}$. For a generic 2-band crossing at a TRI point, the effective Hamiltonian will always be of the form $\sigma\cdot\mathbf{k}$ to leading order, i.e., linear in $\mathbf{k}$. A parabolic band touching is in principle possible but would require additional symmetries, such as rotational symmetry, to suppress the linear order.

For a generic TRI Weyl semimetal, Fermi pockets centered around a TRI momentum can therefore always be associated with a surface Dirac cone.  

\subsection*{H: Generalization to Weyl semimetals with more than four Weyl nodes}

\begin{figure}[t]\centering
\includegraphics[width=0.95\columnwidth]
{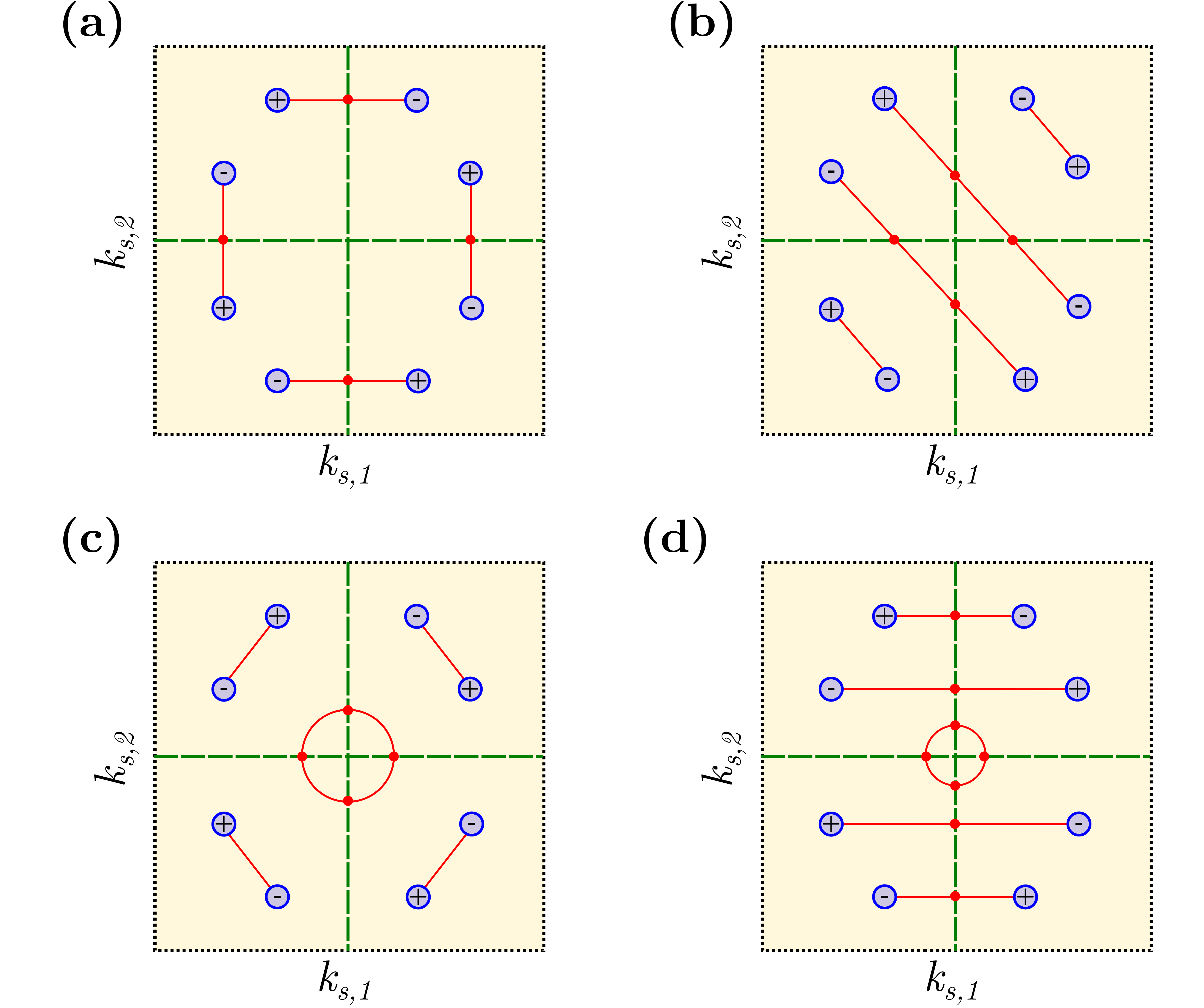}
\caption{(color online) Fermi arc connectivities in the surface BZ of a TRI WSM with eight Weyl points indicated by their topological charge $\pm 1$. The surface projections of 3D TRI planes are highlighted by dotted black ($\nu=0$) or dashed green ($\nu=1$) lines.}
\label{fig:eight_Weyl_nodes}
\end{figure} 

In the main part of the Letter, we presented arguments for the coexistence of Fermi arcs and Dirac cones based on the $\Z_2$ invariants of the Weyl semimetal and on the connectivity of the Weyl nodes. For the sake of simplicity, we illustrated this for a minimal TRI Weyl semimetal with four Weyl nodes. Nevertheless, the scheme can be generalized straight-forwardly to an arbitrary TRI Weyl semimetal with $4n$ Weyl nodes.

For a larger number of Weyl nodes, there are much more possibilities of connecting nodes with opposite charge. This is demonstrated in Fig.~\ref{fig:eight_Weyl_nodes} for a Weyl semimetal with eight Weyl nodes which are all projected to different points in the surface BZ. As for the simple example of four Weyl-node projections, the network of Fermi arcs has to be compatible with the restrictions imposed by the $\Z_2$ invariants of the Weyl semimetal. In our example, the lines $k_{s,1}=0$ and $k_{s,2}=0$ correspond to surface projections of nontrivial TRI planes. Hence, we need to have an odd number of surface Kramers pairs along both of these lines by bulk-boundary correspondence. For the Weyl-node connectivities in Fig.~\ref{fig:eight_Weyl_nodes}(a) and~(b), this is possible without additional Fermi pockets. On the contrary, the connectivities in Fig.~\ref{fig:eight_Weyl_nodes}(c) and~(d) require the presence of an additional Dirac Fermi pocket to be in agreement with the $\Z_2$ invariants. Note that there are \emph{three} surface Kramers pairs along $k_{s,1}=0$ in Fig.~\ref{fig:eight_Weyl_nodes}(d), which is topologically equivalent to the presence of \emph{one} Kramers pair as in the other panels. As before, different Fermi-arc connectivities are connected by Lifshitz transitions which possibly lead to the emergence/disappearance of a Dirac Fermi pocket.

A concrete example for a Weyl semimetal with eight Weyl nodes is given by LaPtBi, which is discussed in the main text and, in more detail, in Sec.~A of this Supplemental Material. For the termination considered there, the bulk Weyl nodes are projected pairwise onto four separate points in the surface BZ giving the projections an effective charge of $\pm 2$. As in the example above, this gives rise to plenty of possible Fermi arc connectivities, some of which require the presence of an additional Dirac Fermi pocket.

\end{document}